\documentclass[fleqn,usenatbib]{mnras}

\usepackage{newtxtext,newtxmath}

\usepackage[T1]{fontenc}

\DeclareRobustCommand{\VAN}[3]{#2}
\let\VANthebibliography\thebibliography
\def\thebibliography{\DeclareRobustCommand{\VAN}[3]{##3}\VANthebibliography}

\usepackage{graphicx}	% Including figure files
\usepackage{amsmath}	% Advanced maths commands
\usepackage{wasysym}
\usepackage{ulem}
\usepackage{soul}

\usepackage[dvipsnames]{xcolor}

\DeclareMathOperator{\sgn}{sgn}

\newcommand*{\bm}[1]{\boldsymbol{\mathbf{#1}}}
\newcommand*{\uv}[1]{\hat{\bm{#1}}}
\newcommand*{\p}[1]{\left(#1\right)}

\title[ZLK+gas]{Gas meets Kozai: the influence of a gas-rich accretion disc on hierarchical triples undergoing von Zeipel-Lidov-Kozai oscillations}

\author[]{
Yubo Su$^{1}$, Connar Rowan$^{2}$ and
Mor Rozner $^{3,4,5}$
\\
$^{1}$ Department of Astrophysical Sciences, Princeton University, 4 Ivy Lane, Princeton, NJ 08540, USA\\
$^{2}$ Niels Bohr International Academy, The Niels Bohr Institute, Blegdamsvej 17, DK-2100, Copenhagen , Denmark \\
$^{3}$Institute of Astronomy, University of Cambridge, Madingley Road, Cambridge CB3 0HA, UK\\
$^{4}$Gonville \& Caius College, Trinity Street, Cambridge, CB2 1TA, UK\\
$^{5}$Institute for Advanced Study, Einstein Drive, Princeton, NJ 08540, USA\\
}

\date{Accepted XXX. Received YYY; in original form ZZZ}

\pubyear{\the\year{}}

\begin{document}
\label{firstpage}
\pagerange{\pageref{firstpage}--\pageref{lastpage}}
\maketitle

% Abstract of the paper
\begin{abstract}
Active galactic nuclei (AGNs) consist of a central supermassive black hole (SMBH) embedded in a region with both high gas and stellar densities:
the gas is present as a thin accretion disc that fuels the central SMBH, while the stars form a dense, roughly isotropic nuclear star cluster.
The binaries present in such a cluster could be considered naturally as triples, with the SMBH as a third object, and their dynamics also depend on the interaction with the gas-rich disc.
In this paper, we study the evolution of such a binary on an inclined orbit with
respect to the disc.
The binary experiences both eccentricity excitation via the von Zeipel-Lidov-Kozai (ZLK) effect and drag forces from each time it penetrates the disc.
We find that, as long as the inner binary remains in the ZLK regime, then the evolution of inner
orbital separation can transition from a regime of gradual hardening to a
regime of rapid softening as the outer orbital inclination decreases.
As such binaries grow wider, their minimum pericentre distances (during ZLK
oscillations) decrease.
We show that a simple geometric
condition, modulated by the complex ZLK evolution, dictates whether a binary expands or contracts due to the interactions
with the AGN disc.
Our results suggest that the interaction with gas-rich accretion disc could
enhance the
rate of stellar mergers and formation of gravitational wave sources, as well as other transients. The treatment introduced here is general and could apply, with the proper modifications, to hierarchical triples in other gas-rich systems.
\end{abstract}

% Select between one and six entries from the list of approved keywords.
% Don't make up new ones.
\begin{keywords}
Accretion discs -- stars:binaries -- galaxies:nuclei
\end{keywords}

%%%%%%%%%%%%%%%%%%%%%%%%%%%%%%%%%%%%%%%%%%%%%%%%%%

%%%%%%%%%%%%%%%%% BODY OF PAPER %%%%%%%%%%%%%%%%%%
\section{Introduction}

Triple systems are ubiquitous over a wide range of scales and astrophysical systems, and they play a key role in various dynamical processes.
About half of Sun-like stars have at least one companion, and the multiplicity fraction increases for higher masses \citep[e.g.,][]{Raghavan2010}.
Triple systems could be roughly divided into two categories: i)
\textit{non-hierarchical} triples, where the relative separation of the three objects is comparable, leading to typically short-lived systems
\citep{Reipurth2000,Stone2019,GinatPerets2021},
and ii) \textit{hierarchical} triples, comprised of a compact binary and distant outer companion, which can be much longer-lived.
Hierarchical triples are abundant in the universe over a wide range of scales and systems, including asteroids \citep{Kozai1962}, planetary binaries \citep[e.g.,][]{Nagasawa2008,PeretsNaoz2009,Naoz2010}, stellar triples \citep[e.g.,][]{Eggleton2001,Duchene2013,Toonen2016}, stellar compact object triples \citep[e.g.,][]{Thompson2011}, black hole binaries in AGN/Galactic nuclei \citep[e.g.,][]{Hoang2018,Fragione2019,Winter2024,Fabj2024} and supermassive black hole systems \citep[e.g.,][]{Blaes2002}.
Their dynamics are rich and are invoked to explain various astrophysical phenomena, including planetary migration \citep[e.g.,][]{Petrovich2015}, tidal disruption events \citep[e.g.,][]{Melchor2024}, and the production of gravitational waves (GWs) sources \citep[e.g.,][]{AntoniniPerets2012,Antognini2014,Hoang2018,Yu2020,su2021_lk90, su2021_massratio, Chandramouli2022, su2024_superthermal, su2025_nsc}.

A key ingredient in the long-term evolution of hierarchical triples is the von Zeipel-Lidov-Kozai effect (ZLK; \citealp{VonZeipel1910,Lidov1962,Kozai1962}; see \citealp{Naoz2016} for a review).
Angular momentum exchange between the inner and outer binaries leads to secular, periodic oscillations in the eccentricity and inclination of the inner binary.
The minimum inclination $i_{\rm crit}$ required between the orbital planes of the inner and outer binaries for ZLK oscillations is
\begin{align}
i_{\rm crit}=\arccos\bigg( \sqrt{\frac{3}{5}}\bigg)\approx39^{\circ}\,.
\end{align}
The ZLK mechanism has been extensively studied in the context of three-body systems \citep[e.g,][]{Caruuba2002,PeretsNaoz2009,AntoniniPerets2012,Shappee2013,Antognini2014,Petrovich2015,Li2015_SMBH_bin,Grishin2016a, GrishinPeretsII2016,
Vanlandingham2016,Stephan2016,Naoz2016,Kimpson2016,Hoang2018,RoznerGrishinPerets2020_PlutoCharon,Yu2020,Britt2021} and has also been discussed in the context of two-body and an additional non-point like mass distributions, e.g.\ an accretion disc \citep[e.g.,][]{chen2013_disk, Martin2014, Fu2015, Fu2015_2, Lubow2017, Suffak2025} and a local star cluster \citep[e.g.,][]{Hamilton2019}.

However, the standard ZLK treatment neglects other environmental effects, such as the presence of gas.
Gas-rich systems are abundant on various scales, including protoplanetary discs, AGN discs, gas-rich star clusters, and the interstellar medium.
The presence of gas can alter the dynamics of stellar, planetary and compact objects, leading to the formation of unique dynamics and transients \citep[e.g.,][]{BahcallOstriker1975,Artymowicz1993,Grishin2016a,
GrishinPeretsII2016,Fan2024,KaurStone2025,Rozner_mAGN2025}
that depend on the geometry and density of the gas distribution, including binary formation \citep[e.g,][]{Li2023,Rowan2023,Rowan2024,DeLaurentiis2023,Rozner2023_bin_form,Whitehead2024,Whitehead2024_novae,Whitehead2025,Dodici2024}, mergers \citep[e.g.,][]{McKernan2012,Stone2017,Yang2019,Tagawa2020,RoznerPerets2022,
Xue2025} and affecting the binary population \citep[e.g.,][]{Mckernan2024_mcfacts,Rowan2024_rates,Rozner_2024_shielding}.
The gravitational effect of a disc of gas on the ZLK effect has also been explored, by considering its modification of the system's secular precession frequencies \citep[e.g.,][]{chen2013_disk, Fan2024}.

In systems such as AGN and protoplanetary discs, the combination of a massive central body and an abundance of gas provides a natural setting to study the dissipative effects of gas on the ZLK oscillations of a binary driven by the central object.
\citet{GrishinPeretsII2016} described ZLK oscillations for planetary systems, fully embedded in a gas-rich protoplanetary disc.
\citet{Lyra2021} describes how nebular gas drag in combination with ZLK oscillations can enhance the probability of planetesimal binary collisions in the solar neighbourhood compared with ZLK alone.
Such effects can also be important for AGN discs, where the gas densities are thought to be very large.
For hierarchical triples in these environments, the ZLK mechanism could be affected significantly and even suppressed through either dynamical or linear hydrodynamic drag. Understanding the significance of gas drag on the ZLK binaries in AGN is vital to interpreting previous studies that indicate that such binaries can lead to black hole mergers \citep[e.g.,][]{AntoniniPerets2012,Vanlandingham2016,Hoang2018}.

In this paper, we use a semi-analytical approach to study the effect of dynamical and linear gas drag from an AGN disc on the orbital dynamics of stellar-mass binaries experiencing ZLK oscillations driven by a central SMBH.
We report on the evolution of both the inner and outer binary orbital elements, focusing on the interplay between ZLK oscillations and gas drag.
The rest of the paper is structured as follows:
We describe the equations of motion in Sec.~\ref{sec:eom} and derive characteristic timescales in Sec.~\ref{sec:gas_drag}. Our results are presented in Sec.~\ref{sec:results}, with their implications discussed in Sec.~\ref{sec:discussion}. We summarise our findings and conclude in Sec.~\ref{sec:summary}.

\section{Equations of Motion}
\label{sec:eom}

\begin{figure}
    \centering
    \includegraphics[width=1\linewidth]{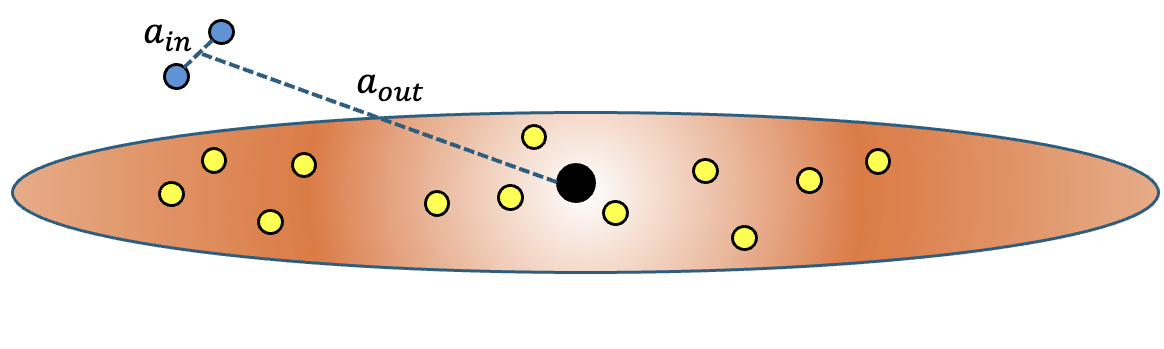}
    \caption{An illustration (not to scale) of the gas-rich accretion disc (in orange), the stars embedded in it (in yellow), the SMBH (in black) and the inner binary companions (in blue). }
    \label{fig:illustration}
\end{figure}

We consider the dynamics of a stellar binary in orbit about a central SMBH in the presence of an AGN disc (see an illustration in Fig. \ref{fig:illustration}).
The inner binary has component masses $m_1$ and $m_2$, radii $R_1$ and $R_2$, semi-major axis $a$, eccentricity $e$, inclination $i$ with respect to the disc normal, and true anomaly $f$.
We generally track its secular orbital evolution by its orbital angular momentum vector $\bm{L} = L\uv{l}$ and its eccentricity vector $\bm{e} = e\uv{e}$.
The two components of the outer binary are the inner binary and the SMBH with mass $m_3$.
Its orbital elements are denoted $a_{\rm out}$, $e_{\rm out}$, $i_{\rm out}$ (also with respect to the disc normal), and its orbit is described by $\bm{L}_{\rm out} = L_{\rm out}\uv{l}_{\rm out}$ and $\bm{e}_{\rm out} = e_{\rm out}\uv{e}_{\rm out}$ with analogous meanings to the inner orbit's properties.
The AGN disc is oriented with its disc normal along $\hat{z}$ and aspect ratio $h \approx 1 / 200$.
We assume that the binary initially satisfies $i_{\rm out} \gg h$, and we aim to model the effect of repeated passages through the AGN disc on the evolution of the hierarchical triple.
Specifically, we consider three effects.
The first is the secular evolution of the hierarchical triple, primarily due to the von Zeipel-Lidov-Kozai (ZLK) effect.
The second is the effect of gas drag on both the inner and outer orbits.
The third is the rapid precession of the outer orbit driven by the gravitational potential of the disc.
We describe each of these in turn below.

\subsection{The von Zeipel-Lidov-Kozai (ZLK) Effect}\label{ss:zlk}

The ZLK effect \citep{VonZeipel1910, Lidov1962, Kozai1962} has been studied extensively in the literature, and we refer the reader to a few excellent studies and resources for a more in-depth discussion \citep[e.g.,][]{liu2015suppression, Naoz2016, shevchenko2016lidov}.
Here, we will briefly describe our approach.

The ZLK effect arises due to the leading-order (quadrupole) mutual gravitational interaction between the inner and outer orbits in a highly misaligned hierarchical triple, and results in large, coupled oscillations in the inner orbit's eccentricity and inclination.
We adopt the standard double-averaged (over the inner and outer orbits), quadrupole-order equations of motion for the ZLK effect \citep[e.g.,][]{liu2015suppression}.
We include three sources of apsidal precession: those due to the tidal and rotational bulges of the binary, and that due to general relativistic periastron advance \citep[e.g.,][]{Kiseleva1998, FabryckyTremain2007}.
A brief recapitulation of these well-known equations is provided in Appendix~\ref{app:eom_zlk}.
For the properties of the star in these expressions, we adopt the stellar mass-radius relations from \citet{tout1996},
and we denote the spin frequencies as a fraction $f$ of the breakup spin rates of the
stars:
\begin{equation}
    \omega_i = f_i (Gm_i/R_i^3)^{1/2},
\end{equation}
where we take $f_i = 0.3$ as a fiducial value.

In order for the double-averaged equations to be accurate, the system is required to satisfy \citep[e.g.,][]{liu2015suppression}
\begin{equation}
    P_{\rm out} \lesssim t_{\rm ZLK}j_{\min},\label{eq:DA_condition}
\end{equation}
where $j_{\min} \equiv \sqrt{1 - e_{\max}^2}$ and $e_{\max}$ is the maximum eccentricity attained by the inner orbit during a ZLK cycle.
$e_{\max}$ depends on both the general-relativistic and rotationally driven apsidal precession via the two dimensionless parameters \citep{liu2015suppression}\footnote{For distant binaries that are not tidally locked, the tidal bulge is much smaller than the rotational bulge for rapidly-rotating bodies \citep[e.g.][]{liu2015suppression}. Note that while $\epsilon_{\rm rot, i}$ is developed for orbit-aligned rotational bulges, the effect of spin-orbit misalignment contributes only an order-unity correction \citep{correia2011_tides}.}:
\begin{align}
% >>> 3 * G * (3 Msun)^2 * (1e4 AU)^3 / ((1 AU)^4 * c^2 * (1e7 Msun))
% 0.026657
    \epsilon_{\rm GR}
        ={}& \frac{3Gm_{12}^2 \tilde{a}_{\rm out}^3}{a_{\rm in}^4c^2m_3}\nonumber\\
        ={}&
            0.03
            \p{\frac{m_{\rm 12}}{3M_\odot}}^2
            \p{\frac{\tilde{a}_{\rm out}}{10^4\;\mathrm{au}}}^3
            \p{\frac{a_{\rm in}}{1\mathrm{au}}}^{-4}
            \p{\frac{m_3}{10^7M_{\odot}}}^{-1}.\label{eq:eps_gr}\\
% >>> (3 Msun) * (1e4 AU)^3 * 0.02 * (Rsun)^2 / ((1 AU)^5 * (1e7 Msun)) * (0.3)^2
% 0.011671
    \epsilon_{\mathrm{rot}, i}
        ={}& \frac{m_{12}\tilde{a}_{\rm out}^3
                k_{2, i}R_i^5}{2Ga^5 m_im_3}\omega_i^2,\\
        ={}& 0.012
            \p{\frac{k_{2, i}}{0.04}}
            \p{\frac{m_{12}}{3M_\odot}}
            \p{\frac{\tilde{a}_{\rm out}}{10^4\;\mathrm{au}}}^3
                \nonumber\\
        &\times \p{\frac{R}{R_{\odot}}}^2
            \p{\frac{a_{\rm in}}{1\;\mathrm{au}}}^{-5}
            \p{\frac{m_3}{10^7M_{\odot}}}^{-1}
            \p{\frac{f_i}{0.3}}^2,
            \label{eq:epsrot}\\
    \frac{9e_{\max}^2}{8}
        ={}&
            \frac{\epsilon_{\rm rot, 1} + \epsilon_{\rm rot, 2}}{3}
                \p{\frac{1}{j_{\min}^3} - 1}
            + \epsilon_{\rm GR}
                \p{\frac{1}{j_{\min}} - 1}
        .
\end{align}
ZLK is truncated when either $\epsilon_{\rm GR} > 9/4$ or $\epsilon_{\rm rot, 1} + \epsilon_{\rm rot, 2} > 9/4$ \citep{liu2015suppression, su2025_nsc}.
For the fiducial parameters, we find that ZLK suppression is dominated by the rotational short-range forces.
This lets us rewrite the condition for double-averaging being valid as
\begin{align}
    P_{\rm out} \lesssim{}& \frac{2t_{\rm ZLK}}{3} (\epsilon_{\rm rot, 1} + \epsilon_{\rm rot, 2})^{1/3}\nonumber\\
        \simeq{}&
            0.2 t_{\rm ZLK}
            \p{\frac{k_{2, i}}{0.04}}^{1/3}
            \p{\frac{m_{12}}{3M_\odot}}^{1/3}
            \p{\frac{\tilde{a}_{\rm out}}{10^4\;\mathrm{au}}}
            \nonumber\\
        &\times \p{\frac{R}{R_{\odot}}}^{2/3}
            \p{\frac{a_{\rm in}}{1\;\mathrm{au}}}^{-5/3}
            \p{\frac{m_3}{10^7M_{\odot}}}^{-1/3}
            \p{\frac{f_i}{0.3}}^{2/3},
\end{align}
easily satisfied in our system, where $t_{\rm ZLK} / P_{\rm out} \approx 2 \times 10^{-3}$.
% >>> ((1e7 Msun) * au^3 / ((3 Msun) * (1e4 AU)^3))^(1/2)
% 0.001826

Finally, our triple must also be dynamically stable.
We adopt the often used condition for dynamical instability from \cite{mardling2001tidal}:
\begin{align}
% >>>  2.8 * (1 + (1e7 / 3))^(2/5)
% 1140.760750
    \frac{a_{\rm out}}{a_{\rm in}}
        &\gtrsim
            2.8\p{1 + \frac{m_3}{m_{12}}}^{2/5}
            \frac{(1 + e_{\rm out})^{2/5}}{(1 - e_{\rm out})^{6/5}}
            \p{1 - 0.3\frac{I_{\rm tot, d}}{180^\circ}}\nonumber\\
        &\sim 10^3
        \p{\frac{m_3}{10^7 M_\odot}}^{2/5}
        \p{\frac{m_{12}}{3 M_\odot}}^{-2/5}
        ,\label{eq:mardling}
\end{align}
where we have taken $I_{\rm out, d} \approx e_{\rm out} \approx 0$ for simplicity, as the circularization and alignment of the outer orbit have comparable-magnitude but opposite effects on the stability condition.

\subsection{Effect of Gas on Binary Orbit}

There are several models for the rate that an object decelerates when moving through a gaseous medium, including gas dynamical friction \citep[e.g.,][]{Ostriker1999,ONeill2024} and disc-like migration/formation of mini-discs \citep[e.g.,][]{Artymowicz1991,McKernan2012,Stone2017,Tagawa2020}.
In this work, we adopt the so-called ``aerodynamic'' drag prescription \citep[e.g.,][]{Artymowicz1993, SubsrKaras1999,
GenerozovPerets2022, wang2024_agn} given by
\begin{equation}
    \bm{F}_{\rm d}(\bm{v}_{\rm rel})
        =
            -\frac{\mathcal{C}_D}{2}
                A
                \rho_g
                v_{\rm rel}
                \bm{v}_{\rm rel}.\label{eq:def_Fdrag}
\end{equation}
Here, the relative velocities for the individual masses is $\bm{v}_{\rm rel, i} \equiv \bm{v}_{\rm out} - \bm{v}_{\rm disc} + \bm{v}_{i}$ where $\bm{v}_{\rm out}$, $\bm{v}_{\rm disc}$ and $\bm{v}_{i}$ are the outer binary velocity, disc Keplerian velocity and individual inner binary object velocity respectively.
The cross section $A$ is given by the sum\footnote{We adopt the sum rather than the maximum as done in previous works \citep[e.g.,][]{wang2024_agn} to ensure continuity of the drag force as the orbit evolves.} of the object's geometric radius and its gravitational radius of influence (corresponding to the characteristic radius of Bondi-Hoyle-Lyttleton accretion, \citealp{hoylelyttleton_1939, bondi1944_BHL, edgar2004_BHLreview}), i.e.
\begin{equation}
    A \equiv \pi R^2 + \pi\p{\frac{Gm}{v_{\rm rel}^2 + c_{\rm s}^2}}^2,\label{eq:def_A_drag}
\end{equation}
where $R$ is the star's physical radius and $c_{\rm s}$ is the local sound speed \citep[e.g.,\ a continuous version of the prescription adopted in][]{wang2024_agn}\footnote{
Note that $A \ll \pi r_{\rm H}^2$ for binaries crossing the disk on such inclined trajectories, where $r_{\rm H} = a_\text{out}(m_{12}/(3m_3))^{1/3}$ is the Hill radius.
As such, the scale of the accretion flow is set by the BHL cross section $A$, within which the SMBH gravity is negligible.}.
For a sense of scale, we comment that
\begin{align}
    \frac{A_{\rm phys}}{A_{\rm grav}}
        &\equiv
            \p{\frac{v_{\rm rel}^2R}{Gm}}^2
        \simeq
            \p{\frac{v_{\rm out}^2\sin^2 i_{\rm out} R}{Gm}}^2\nonumber\\
        &\approx
% >>> 1 / (G * Msun / (G * (1e7 Msun) / (1e4 AU) * Rsun))^2
% 21.613798
            20\sin^2i_{\rm out}
            \p{\frac{M_{\rm SMBH} / m}{10^{7}}}^2
            \p{\frac{R / a_{\rm out}}{R_{\odot} / 10^4\;\mathrm{au}}}^2
            ,\label{eq:crosssection_ratio}
\end{align}
so the drag due to the physical and gravitational cross-section of the objects
is comparable for a typical binary in an AGN\@.
For convenience, we will denote the $A_{\rm phys} \gg A_{\rm grav}$ to be the geometric regime, and the opposite to be the BHL (Bondi-Hoyle-Lyttleton) regime.
The disc velocity is assumed to be Keplerian.
The evolution of both the inner and outer binaries can then be obtained by expressing the drag experienced by the binary's centre of mass and reduced mass in terms of the drags exerted on each binary component:
\begin{equation}\begin{aligned}
    \bm{F}_{\rm d, com}
        &= \bm{F}_{\rm d, 1} + \bm{F}_{\rm d, 2},\\
    \bm{F}_{\rm d, red}
        &= \mu
            \left(\frac{\bm{F}_{\rm d, 2}}{m_2} - \frac{\bm{F}_{\rm d, 1}}{m_1}\right).\label{eq:F_d_decomposition}
\end{aligned}\end{equation}
The former is responsible for changes to the outer orbit, and the latter for changes to the inner orbit.
These forces can be used to conveniently express the changes to the inner and outer orbits via the expressions provided in Appendix~\ref{app:eom_disc}.

In addition to these above results, note that each disc passage occurs over a timescale comparable to the inner orbital period:
\begin{align}
    t_{\rm cross} &\simeq
            185\;\mathrm{day} \frac{1}{\sin i_{\rm out}}
            \left(\frac{a_{\rm out}}{10^4\;\mathrm{au}}\right)^{1/2}
            \left(\frac{h}{0.005}\right),\\
    P_{\rm in}
        &=
            200\;\mathrm{day}
            \left(\frac{m_{12}}{3 M_\odot}\right)^{-1/2}
            \left(\frac{a_{\rm in}}{\mathrm{au}}\right)^{3/2}.
\end{align}
% >>> 2 * pi / (G * (3 Msun) / (1 AU)^3)^(1/2) / day
% 210.859123
% >>> 2 * (50 AU) / (G * (1e7 Msun) / (1e4 AU))^(1/2) / day
% 183.811701
As such, we evolve the inner orbit according to its time-averaged rate of change over an inner orbital period.
% over its true anomaly $f$ (using that $\dot{f} = l / r^2$, where $l \equiv
% \sqrt{Gm_{12}a(1-e^2)}$ is the inner binary's specific angular momentum).

\subsection{Disc-Driven Precession}\label{ss:diskp_eom}

Finally, due to the large mass of the disc, it will drive orbital precession of the outer orbit; the effect on the inner orbit is weaker by a factor of $(a / a_{\rm out})^2$ and is negligible.
While this has been (reasonably) omitted in previous studies of interactions of single stars with AGN discs, it cannot be neglected here, since the outer orbit orientation plays a key role in the ZLK dynamics of the inner binary.
There are two classes of expressions for this precession effect available in the literature, but neither are appropriate here:
some \citep[e.g.,][]{terquem2010_disk, zhao2012_inclineddiskprec, chen2013_disk} are derived in the limit where $a_{\rm out} \ll r_{\rm in}$ (i.e.,\ where the disc is exterior to the precessing orbit), while others \citep[e.g.][]{ward1981, hahn2003, zanazzilai2018} assume that the orbital inclination is smaller than the disk aspect ratio (i.e.,\ that the orbit is fully embedded in the disk).
% In the case where the orbit and a thin disc have comparable radii, the precession rate is formally divergent: multipolar expansions break down near $a_{\rm out} \simeq r$, and the Laplace coefficients in standard Laplace-Lagrange theory are also singular.
Instead, we show in Appendix~\ref{app:prec} that the precession induced by a disc on an orbit with semi-major axis $a_{\rm out}$ depends on the gas density $\rho_g$ of the disc at $a_{\rm out}$ via:
\begin{align}
    \frac{d\bm{L}_{\rm out}}{dt}
        \simeq{}&
            \Omega_{\rm out}
                (\bm{L}_{\rm out} \times \uv{z}),\\
    \Omega_{\rm out}
        \simeq{}&
            \frac{G \rho_g}{n_{\rm out}}\frac{4h}{\sin i_{\rm out}}\nonumber\\
% >>> 2 * pi / (G * (1e11 Msun) / (1 pc)^3 / (G * (1e7 Msun) / (1e4 au)^3)^(1/2) * 0.005 * 4) / yr
% 1.387295e+04
        ={}&
            \frac{2\pi}{10\;\mathrm{kyr}}
            \frac{1}{\sin i_{\rm out}}
            \left(\frac{\rho_g}{10^{11}\;\mathrm{M_\odot / pc^3}}\right)
            \left(\frac{h}{0.005}\right)\nonumber\\
            &\times
                \left(\frac{m_3}{10^7 M_{\odot}}\right)^{-1/2}
                \left(\frac{a_{\rm out}}{10^4\;\mathrm{au}}\right)^{3/2},
\end{align}
and identically for $\bm{e}_{\rm out}$, where $h \equiv H / r$ is the disk aspect ratio.
Especially as the outer orbit begins to align with the disc, this is substantially faster than the next fastest timescale in the system.
We adopt a fixed value of $\Omega_{\rm out} = 2\pi / 10\;\mathrm{kyr}$ for computational efficiency, as further increase of $\Omega_{\rm out}$ will not affect the dynamics.

\section{Secular dynamics in gas-rich environments}
\label{sec:gas_drag}

In this section, we will discuss some unique effects that are expected to occur for secular evolution in gas-rich environments.

\subsection{Characteristic timescales}

To estimate the gas densities at which gas drag can fully suppress ZLK oscillations, we can compare the secular and drag torques on the inner orbit, denoted $T$.
Note that the effect of gas drag on the inner orbit must be evaluated by considering the orbit-averaged effect of the gas drag on \textit{only} the reduced mass of the inner binary (see later discussion in Section~\ref{sec:results}, notably Eqs.~\ref{eq:edot_geom} and~\ref{eq:edot_grav}).
This gives
\begin{align}
    T_{\rm drag}
        \sim{}&
            aF_{\rm d} \sim a\rho_g A v_{\rm in}v_{\rm out},\nonumber\\
    T_{\rm ZLK}
        \sim{}&
            \frac{L}{t_{\rm ZLK}}
            \sim \mu a^2 \frac{Gm_{123}}{a_{\rm out}^3},\nonumber\\
    \frac{T_{\rm drag}}{T_{\rm ZLK}}
        ={}&
            \frac{\rho_g Aa_{\rm out}^{2.5}}{\mu
                a^{1.5}}\sqrt{\frac{m_{12}}{m_{123}}}\nonumber\\
% >>> ((1e11 Msun) / pc^3) * (pi * Rsun^2) * (1e4 au)^(2.5) / ((0.25 Msun) * au^(1.5)) * (3 / 1e7)^(1/2)
% 0.016954
        \approx{}&
            0.017
                \left(\frac{\rho_g}{10^{11} M_{\odot} / \mathrm{pc}^3}\right)
                \left(\frac{a_{\rm out}}{10^4\;\mathrm{au}}\right)^{2.5}
                \left(\frac{a}{1\;\mathrm{au}}\right)^{-1.5},\label{eq:dragtorque_ratio}
\end{align}
where we have evaluated the drag using the binary's geometric cross section.
Thus, we see that gas drag plays a modestly subdominant role within each ZLK cycle.

We also note that the outer orbit evolves on a characteristic timescale (due to drag)
\begin{align}
    t_{\rm out, drag}
        \sim{}& \frac{L_{\rm out}}{a_{\rm out} F_{\rm d, com}}
        \sim
            \frac{\mu_{\rm out} a_{\rm out} v_{\rm out}}
            {a_{\rm out} \rho_g A v_{\rm out}^2}\nonumber\\
        ={}&
            20\;\mathrm{Myr}
            \left(\frac{\rho_g}{10^{11} M_{\odot} / \mathrm{pc}^3}\right)^{-1}
                \left(\frac{a_{\rm out}}{10^4\;\mathrm{au}}\right)^{1/2}
                \p{\frac{\mu_{\rm out}}{3M_{\odot}}}.\label{eq:t_outdrag}
\end{align}
% >>> (3 Msun) / ((1e11 Msun) / (1 pc)^3 * pi * Rsun^2 * (G * (1e7 Msun) / (1e4 AU))^(1/2)) / yr
% 1.951009e+07
This is in agreement with results obtained by other authors \citep[e.g.,][]{GenerozovPerets2022}, who find $\sim 50\%$ of single objects can be successfully realigned in the $\sim 10\;\mathrm{Myr}$ lifetime of an AGN disc with comparable gas densities.

By contrast, the inner orbit evolves on characteristic timescale
\begin{align}
    t_{\rm in, drag}
        \sim{}& \frac{L}{aF_{\rm d, red}}
        \sim
            \frac{\mu a v}
            {a \rho_g A vv_{\rm out}}\nonumber\\
        ={}& t_{\rm out, drag} \frac{\mu}{m_{12}}.\label{eq:t_indrag}
\end{align}
This shows that the inner binary evolves on a \textit{comparable} time scale to the outer binary, differing by a factor of $q/(1+q)^2$ with $q = m_2 / m_1 < 1$ the inner binary mass ratio.

Note that we do not take into account any transitory "switching on" phase of the AGN and assume a static disc. It is suggested that AGN accrete episodically with growth phases of $\sim10^{5} \ \rm{yr}$ \citep[e.g.,][]{Schawinski2015}, so this ramp-up phase is assumed to be small compared to the lifetime of an accretion phase and the binary immediately sees a fully established AGN disc.

\subsection{A more realistic disc model}

\begin{figure*}
    \centering
    \includegraphics[width=17cm]{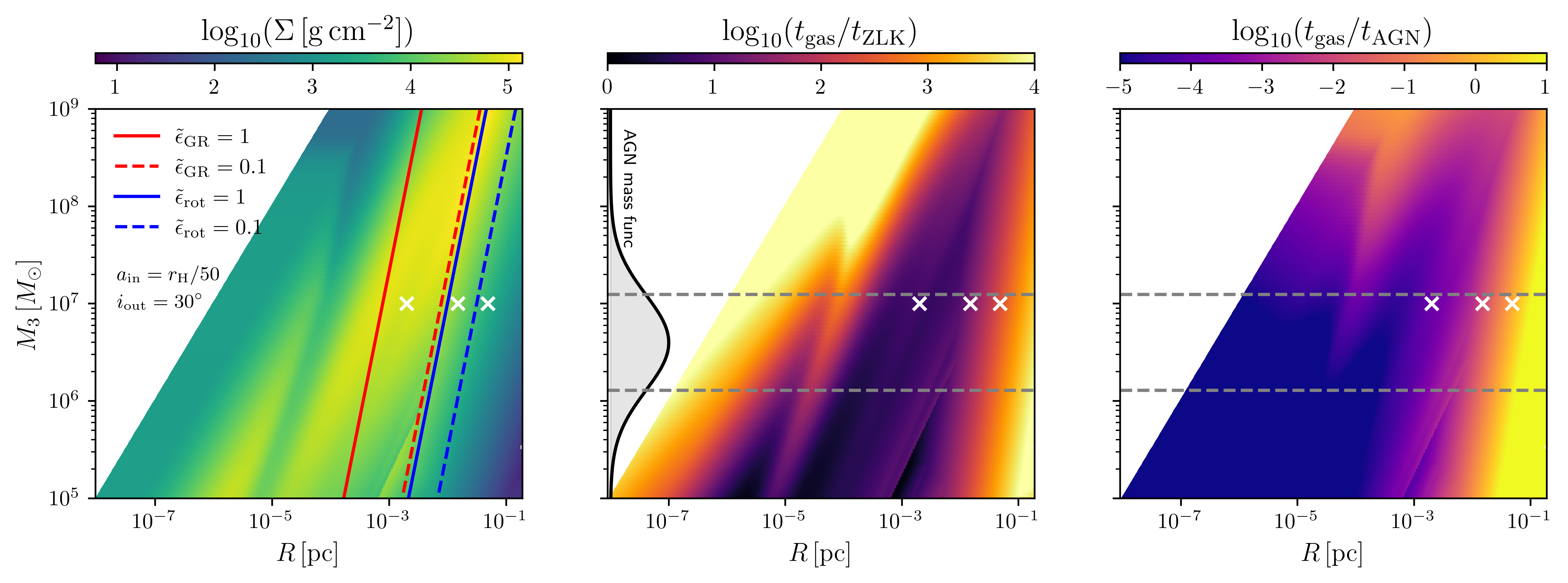}
    \caption{Visualization of the AGN parameter space in distance $R$ from the SMBH and SMBH mass $M_3$ where ZLK oscillations and gas drag become important considerations in the evolution of satellite binaries. We maintain our fiducial masses of $m_1$, $m_2 = 2M_\odot$, $M_\odot$ and assume that $i_\text{out}=30^\circ$, $e_\text{out}=0$. \textit{Left:} the AGN surface density, the red and blue lines mark the maximal radii wherei.e.(general relativity) precession (Eq. \ref{eq:eps_gr}) and stellar rotation (Eq. \ref{eq:epsrot}) induced precession will respectively dominate over ZLK oscillations and we define $\tilde{\epsilon}=\frac{4}{9}\epsilon$. \textit{Middle:} The ratio of the gas drag (Eq. \ref{eq:t_indrag}) to ZLK timescale (Eq. \ref{eq:t_ZLK}), also shown is the AGN mass function and $1-\sigma$ deviation (dashed lines). \textit{Right:} the ratio of the gas drag timescale to AGN timescale $t_\text{AGN}$. The disc properties were created using \texttt{pAGN} assuming $\alpha=0.1$, $X=0.7$, $\epsilon_\text{eff}=0.1$ and $L_\text{E}=0.1$. The markers indicate where our three realistic example simulations lie in the parameter space, see Sec.~\ref{ss:sim_realistic}.}
    \label{fig:connar}
\end{figure*}

In Fig.~\ref{fig:connar}, we illustrate the hierarchy of these timescales using a more realistic AGN disc profile. Using the purpose build \texttt{python} package \texttt{pagn} \citep[e.g][]{Gangardt2024} we generate 200 models spanning SMBH masses in the range $m_3= 10^{5}M_\odot$ to $10^{9}M_\odot$. The disc models assume a Sirko-Goodman \citep{sirko2003_agn} $\alpha$-disc with $\alpha=0.1$, a SMBH with an Eddington fraction $L_\mathrm{E}=0.1$, radiative efficiency $\epsilon_\text{eff}=0.1$ and hydrogen/helium fractions $X/Y = 0.7/0.3$. We also show the observed AGN SMBH mass function according to \citet{Greene_Ho2007,Greene2009} to indicate the parameters associated with a "typical" AGN.

The ratio of the gaseous to ZLK timescale serves as a proxy for the level of influence we can expect gas effects to alter the ZLK evolution of the inner binary. Unsurprisingly, this closely follows the $\Sigma$ distribution. Considering binaries of equal hardness, i.e $a_\text{in}/r_\text{H}$ is constant, where $r_\text{H}$ is again the Hill radius, then under the assumption that comparatively few binaries will exist within $R<10^{-4} \ \rm{pc}$ we would expect binaries undergoing ZLK oscillations to be increasingly affected by gaseous drag. We find this result is largely independent of the choice of $m_3$, with $t_\text{gas}/t_\text{ZLK}$ ranging from $10^3$ to $<10^{1}$ for $R= 10^{-1} \ \rm{pc}$ to $10^{-3} \ \rm{pc}$ for typical AGN SMBH masses.

The ZLK mechanism can be suppressed at increasingly large $a_\text{out}$ for larger SMBH masses. For our fiducial rotation of $f_i=0.3$, rotation suppression is dominant to GR effects. The suppression condition of $\epsilon_\text{rot}>9/4$ is met at $R\sim10^{-3} \ \rm{pc}$ and $\sim10^{-2} \ \rm{pc}$ for low and high $m_3$ respectively.

One can also compare the gas timescale to the AGN timescale $t_\text{AGN}$. For gas to significantly affect the inner binary orbit during the lifetime of the AGN disc, we require $t_\text{gas}/t_\text{AGN}\leq1$, which is satisfied for $R<0.1 \ \rm{pc}$ for a typical AGN. Putting this together, we expect the combined gas-ZLK regime to exist on the scales of $10^{-3}-10^{-1} \ \rm{pc}$ with gas playing an increasingly significant role towards the lower end of this range.

\section{Results}\label{sec:results}

\begin{figure*}
    \centering
    \includegraphics[width=\textwidth]{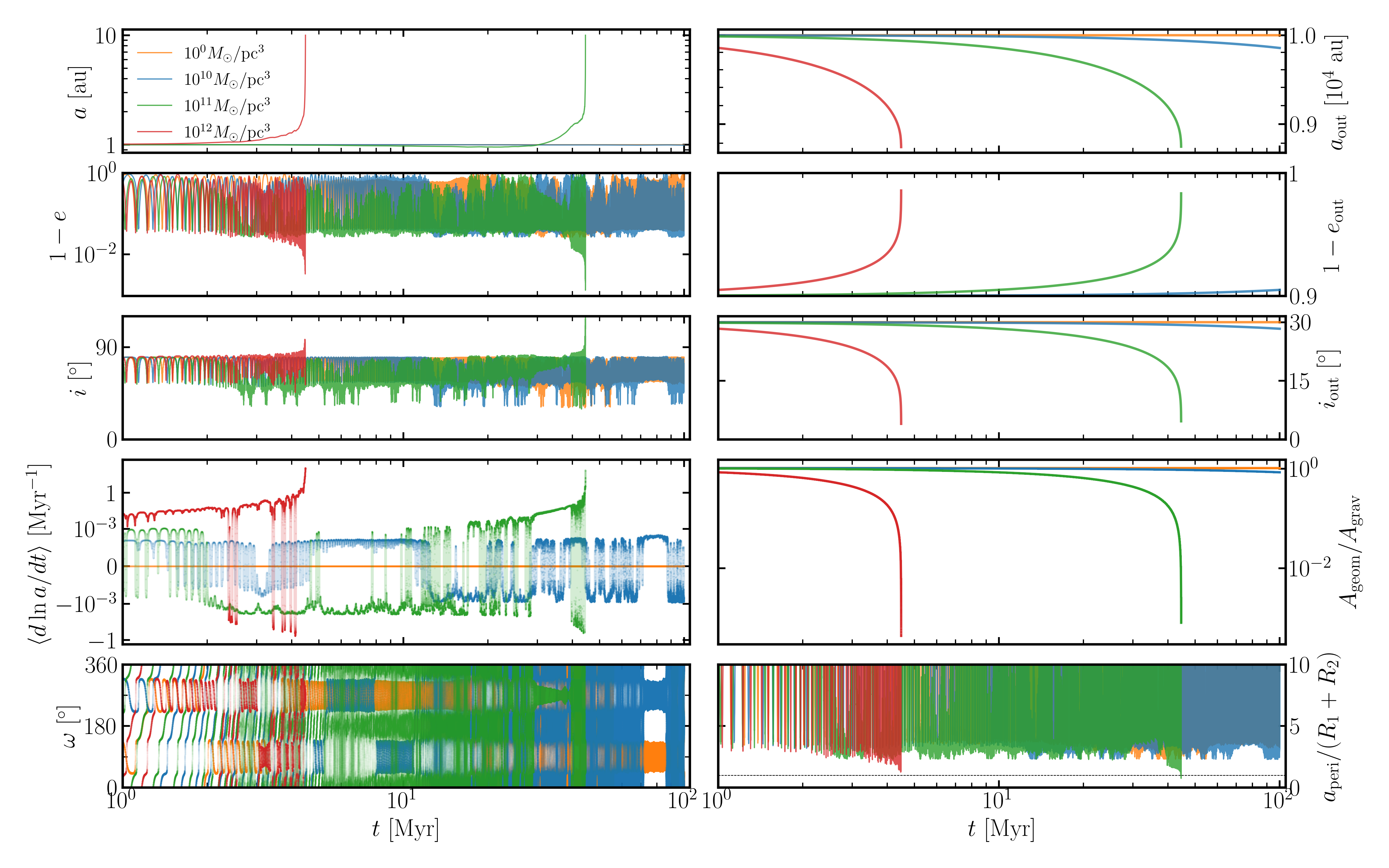}
    \caption{
    Evolution of an unequal mass-ratio system ($m_1=2M_{\odot}$ and $m_2 =
    M_{\odot}$) with both stars rotating at $\omega_i = 0.3\sqrt{Gm_i/R_i^3}$
    for four different gas densities (the value in the legend corresponds to the density at $a_{\rm out}$, and we take $\Sigma(r) = \Sigma(a_{\rm out}) a_{\rm out} / r$); we use $a = 1\;\mathrm{au}$, $a_{\rm out}
    = 10^4\;\mathrm{au}$, $e_{\rm out} = 0.1$, $i_{\rm out} = 30^\circ$, and we
    take initial values $e_0 = 0.2$ and $i_0 = 80^\circ$.
    The left column of panels shows the evolution of the inner orbit:
    the top panel illustrates that the orbit eventually \textit{softens},
    particularly at late times.
    The second and third panels show the signature ZLK eccentricity-inclination
    cycles, though the inclination is given with respect to $\uv{z}$, the disc
    normal.
    The stochasticity in the eccentricity and
    inclination cycles is due to the slight aperiodicity introduced by
    the finite precession rate of the outer orbit driven by the disc.
    The fourth panel shows the instantaneous $\dot{a}$, where significant
    structure can be seen; see detailed discussion in Section~\ref{ss:ain_dot}.
    The final panel shows the inner orbit's argument of pericentre $\omega$ with respect to the disc plane.
    The first three panels of the right column show the evolution of the outer
    orbit: the orbit shrinks, circularizes, and aligns with the disc plane.
    The fourth panel shows the ratio of the geometric and BHL
    contributions to the drag force cross-section
    (Eq.~\ref{eq:crosssection_ratio}).
    The fifth panel shows the ratio of the orbit's pericentre distance to the radii of the stars; recall that the Roche radius of a star is $\sim 2R_\star$.
    }
    \label{fig:results}
\end{figure*}

We consider a binary with $m_1 = 2M_\odot$ and $m_2 = M_\odot$ in orbit around a SMBH with $m_3 = 10^7 M_\odot$.
For ease of comparison, we take the separation of the inner orbit to be $a = 1\;\mathrm{au}$ and of the outer orbit to be $a_{\rm out} = 10^4\;\mathrm{au}$, but we allow the disc surface density to be a free parameter.
We truncate the system evolution once $i_{\rm out}$ decreases below $h = 1/200$, whereupon the binary becomes fully embedded in the disc.
The results of such an evolution are shown for four gas surface densities in
Fig.~\ref{fig:results}, where the disc surface density is reported at $a_{\rm out}$; we take $\Sigma = \Sigma(a_{\rm out}) (a_{\rm out} / r_{\rm out})$, i.e.\ a $p = -1$ surface density profile.
We see that the outer orbit evolves in the expected manner, where the orbit
shrinks, circularizes, and aligns with the disc \citep[e.g.,][]{wang2024_agn}.

\begin{figure*}
    \centering
    \includegraphics[width=\textwidth]{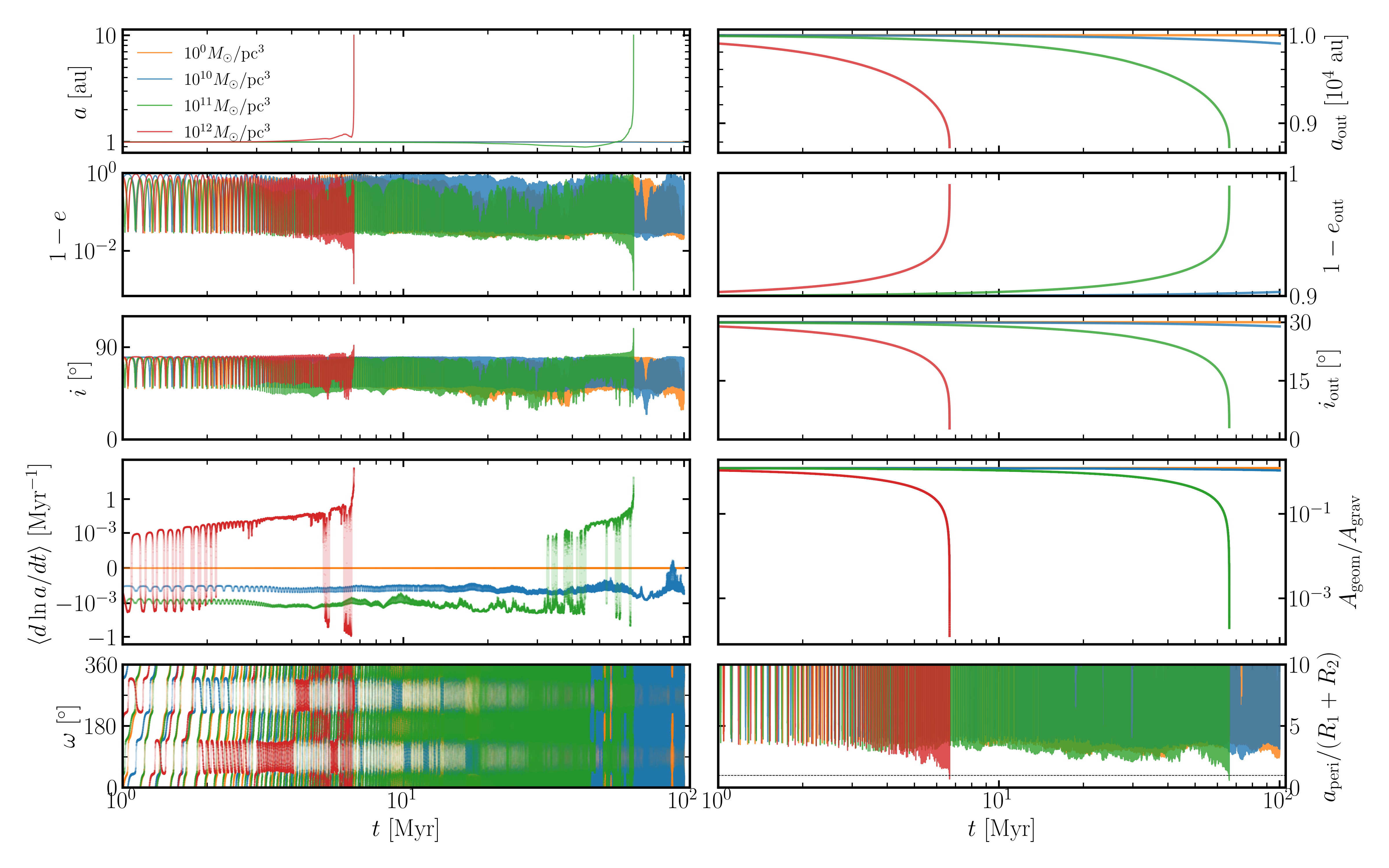}
    \caption{
    Same as Fig.~\ref{fig:results} but for equal-mass binary, $m_1 = m_2 = M_{\odot}$.
    The evolution is qualitatively similar.
    }\label{fig:results2}
\end{figure*}

On the other hand, the inner orbit undergoes several phenomena worth discussing
in detail.
First, note that the inner orbit executes ZLK oscillations about $\hat{z}$
instead of $\hat{l}_{\rm out}$ (bottom-right panels of Fig.~\ref{fig:results}).
This is because $\hat{l}_{\rm out}$ is rapidly varying, with average orientation
along $\hat{z}$.
Second, note that the inner orbit's semi-major axis evolution broadly
transitions from inspiral to outspiral as the outer orbital inclination
decreases.
While the outspiraling regime has been previously characterized
\citep{sanchezsalcedo2014_gasdf, GrishinPeretsII2016}, its dependence on the outer
orbit's geometry (when in the ZLK regime) is novel
to this work.
For comparison, we also display the evolution of an equal mass ratio binary in
Fig.~\ref{fig:results2}, with all other parameters held constant; no significant
qualitative changes are seen in the evolution.

\subsection{Inner Orbital Evolution: Transition between Inspiral and
Outspiral}\label{ss:ain_dot}

To better understand the evolution of the semi-major axis, we display $\dot{a}$, slightly time-averaged with a sliding Gaussian window with width
$3\;\mathrm{kyr}$, in the bottom-left panel of Fig.~\ref{fig:results}.
The overall trend and detailed structure of this panel are the subject of this section.

Without loss of generality, we consider just one of the two disc passages; the same conclusion holds for both disc passages.
In addition, we focus our discussion on
the case of circular orbits since, during ZLK cycles, most time is spent at low
eccentricities.
The instantaneous energy transfer rate into the inner orbit, which is related to the evolution of the semi-major axis by $E \propto -1/a_{\rm in}$, which has orbital velocity $\bm{v}$, is given by
\begin{align}
    \dot{E}
        ={}& \bm{F}_{\rm d, red} \cdot \bm{v}.
\end{align}
Now, note that the drag force is typically dominated by the relative velocity of the outer
orbit (since $|\bm{v}_{\rm out, rel}| = |\bm{v}_{\rm out} - \bm{v}_{\rm disc}|
\gg v$).
So we can linearly expand $\bm{F}_{\rm d, red}$ as:
\begin{align}
    \bm{F}_{\rm d, red}
        ={}&
            \mu\p{\frac{\bm{F}_{\rm d, 2}}{m_2}
                - \frac{\bm{F}_{\rm d, 1}}{m_1}}\nonumber\\
        \approx{}&
            \mu\Bigg(\frac{\bm{F}_{\rm d, 2}\p{\bm{v}_{\rm out, rel}}
                    + \delta \bm{F}_{\rm d, 2}\p{\bm{v}_{\rm
                    2}}}{m_2}\nonumber\\
            &- \frac{\bm{F}_{\rm d, 1}\p{\bm{v}_{\rm out, rel}}
                    + \delta \bm{F}_{\rm d, 1}\p{\bm{v}_{\rm 1}}}{m_1}\Bigg),
\end{align}
where $\delta$ denotes the first perturbation.
This can be rewritten by defining the symmetric drag
\begin{equation}
    \tilde{\bm{F}}_{\rm d}(\bm{v}_{\rm rel})
        \equiv
            -\frac{C_D}{2}\rho_g (A_1 + A_2)
                v_{\rm rel}\bm{v}_{\rm rel}.
\end{equation}
The linear expansion then becomes \citep[cf. Eq.~12 of][]{GrishinPeretsII2016}:
\begin{equation}
    \bm{F}_{\rm d, red}
        =
            \tilde{\bm{F}}_{\rm d}\p{\bm{v}_{\rm out, rel}}
            \frac{m_1 - m_2}{m_1 + m_2}
            +
            \delta \tilde{\bm{F}}_{\rm d}\p{\bm{v}},
\end{equation}

We can then directly evaluate the perturbative piece for the two limits of our
drag prescription.
In the geometric limit ($A_{\rm geom} \gg A_{\rm grav}$), the result is
\begin{align}
    \frac{2}{C_D\pi (R_1^2 + R_2^2)\rho_g}\dot{E}
        % \approx{}&
        %     -
        %     [v_{\rm out, rel}\bm{v}_{\rm out, rel}\frac{m_1 - m_2}{m_1 + m_2}\nonumber\\
        %     &+ (\bm{v} \cdot \hat{\bm{v}}_{\rm out, rel})\bm{v}_{\rm out, rel}
        %     + v_{\rm out, rel} \bm{v}] \cdot \bm{v}\nonumber\\
        \approx{}&
            \left\{-v_{\rm out, rel}\bm{v} \cdot \bm{v}_{\rm out, rel}\frac{m_1 - m_2}{m_1 + m_2}\right\}\nonumber\\
            &-[v_{\rm out, rel}(\hat{\bm{v}}_{\rm out, rel}\cdot \bm{v})^2
                + v_{\rm out, rel} v^2],\label{eq:edot_geom}
\end{align}
On the other hand, in the BHL limit (we assume $c_s \ll v_{\rm rel}$, which will be valid for $i_{\rm out} \gg H / r$)
\begin{align}
    \frac{2}{C_D\pi G^2(M_1^2 + M_2^2)\rho_g}\dot{E}
        % \approx{}&
        %     -
        %     \Big[\frac{\bm{v}_{\rm out, rel}}{v_{\rm out, rel}^3}\frac{m_1 - m_2}{m_1 + m_2} + \frac{\bm{v}}{v_{\rm out}^3}\nonumber\\
        %         &- \frac{\bm{v}_{\rm out, rel}}{v_{\rm out}^4}
        %         3\bm{v} \cdot \hat{\bm{v}}_{\rm out, rel}\Big]
        %         \cdot \bm{v}\nonumber\\
        \approx{}&
            \left\{-\frac{\bm{v} \cdot \bm{v}_{\rm out, rel}}{v_{\rm out, rel}^3}
                \frac{m_1 - m_2}{m_1 + m_2}\right\}\nonumber\\
            &- \frac{v^2 - 3(\bm{v} \cdot \hat{\bm{v}}_{\rm out, rel})^2}{v_{\rm out}^3},\label{eq:edot_grav}
\end{align}

First, we note that the terms in Eqs.~\eqref{eq:edot_geom}
and~\eqref{eq:edot_grav} that are $\propto v^1$ (the terms surrounded in curly braces) will vanish when averaged over the inner orbit.
As such, the $\dot{E}$ is independent of the inner binary's mass ratio.

\begin{figure*}
    \centering
    \includegraphics[width=\textwidth]{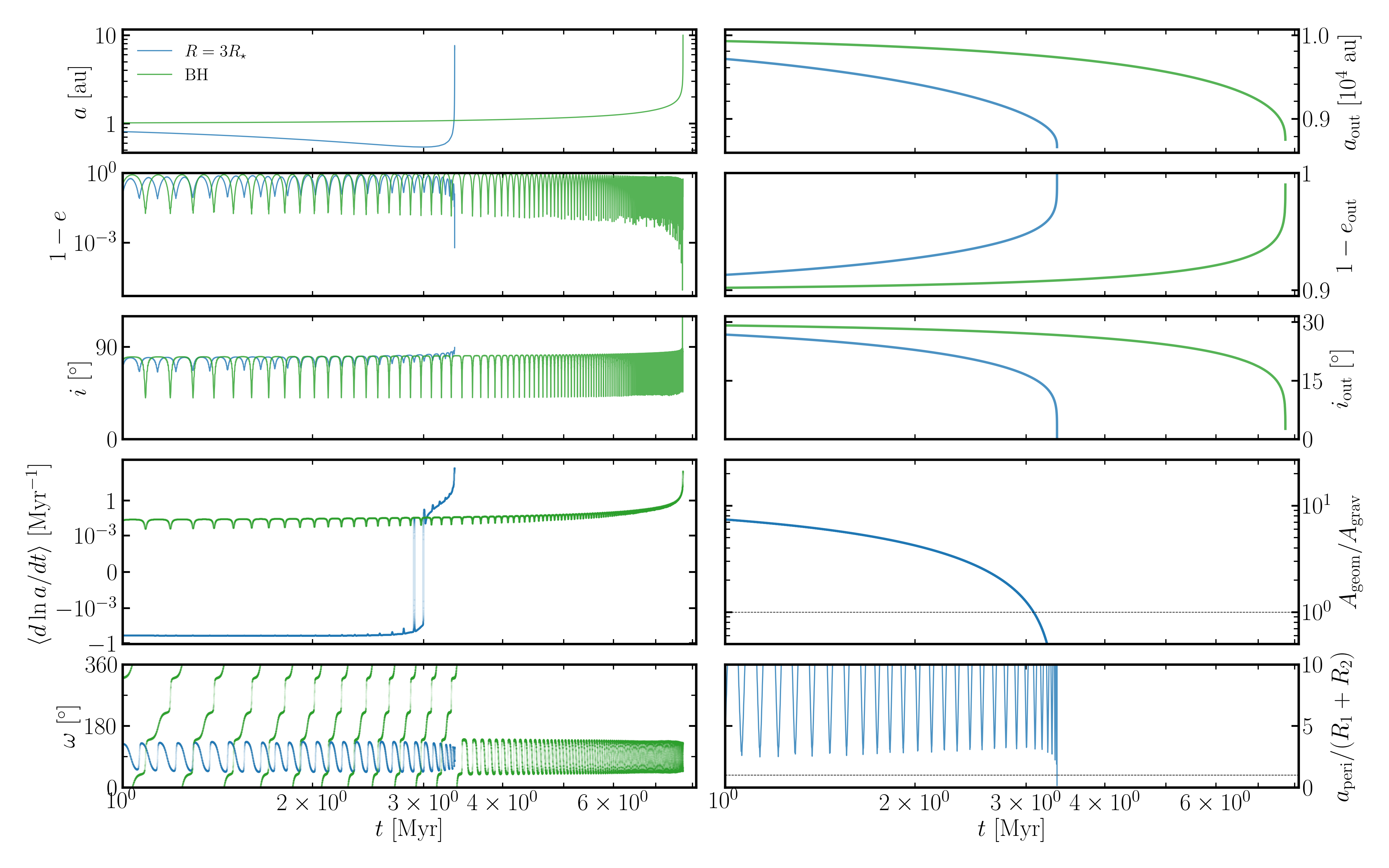}
    \caption{
    Same as Fig.~\ref{fig:results2} but comparing a stellar binary with its radii increased by a factor of
    $3\times$ to a black hole binary ($R = 0$), for which $A_{\rm geom} = 0$.
    The dependence of the direction of orbital evolution (hardening or softening) on the regime of drag is clearly evident by comparing the fourth row of panels.
    }\label{fig:results_puffy}
\end{figure*}

From the remaining $\propto v^2$ terms,
we see from Eq.~\eqref{eq:edot_geom} that inspiral should always occur in the
geometric drag regime.
This is corroborated by the blue lines shown in Fig.~\ref{fig:results_puffy}, which
shows the evolution of a stellar binary with artificially inflated ($3\times$)
radii\footnote{Physically, this is analogous to a decrease in $a_{\rm out}$
(Eq.~\ref{eq:crosssection_ratio}), but we have chosen to directly modify $R$ so
that the ZLK evolution is unaffected, facilitating easier comparison among
evolutionary tracks.}.
On the other hand, for the BHL drag regime, we see that both in and outspiral
can occur.
This can be seen from the green lines Fig.~\ref{fig:results_puffy}, where we show that a binary black
hole undergoing ZLK experiences consistent orbital expansion.

\begin{figure}
    \centering
    \includegraphics[width=\columnwidth]{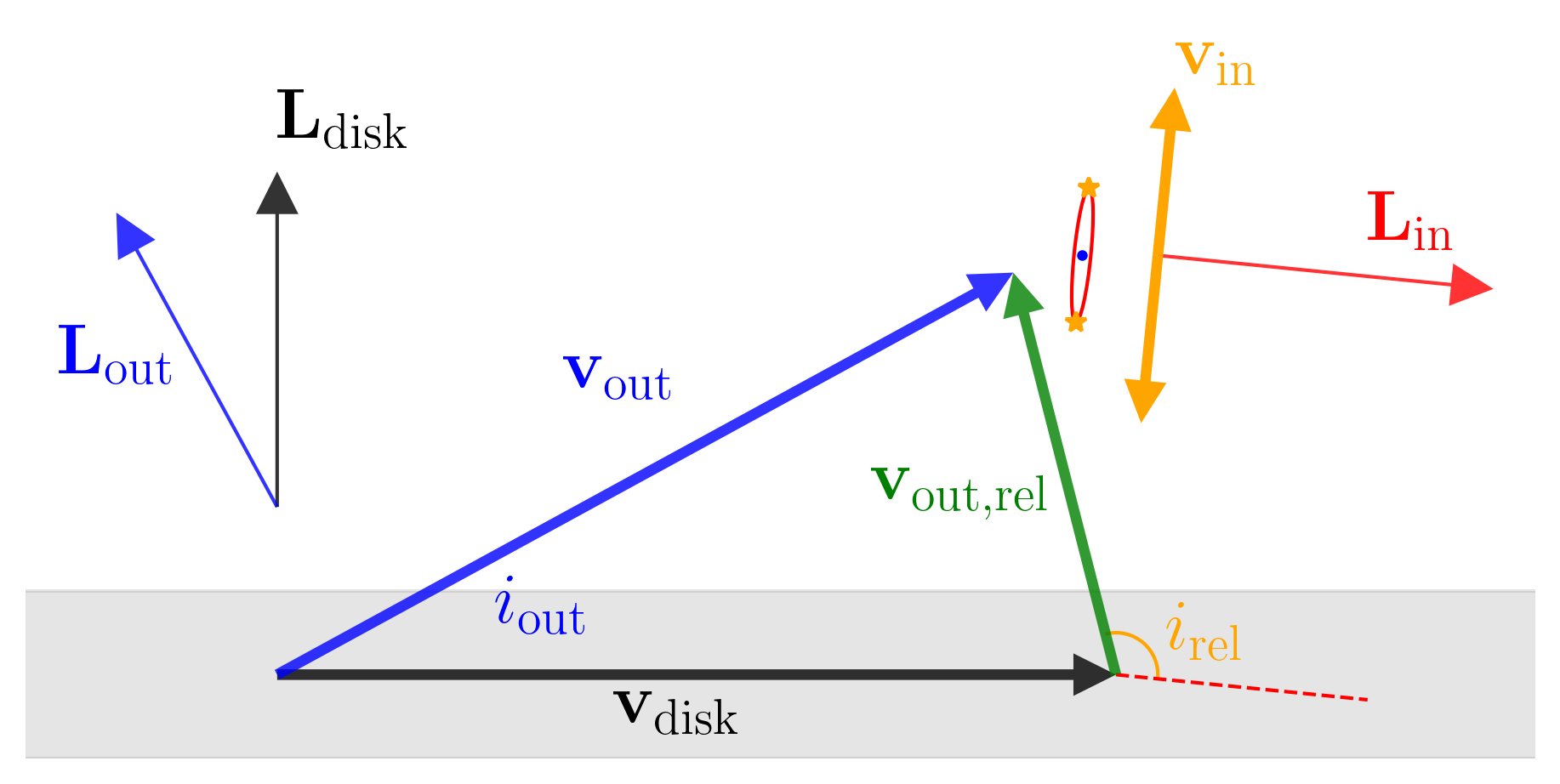}
    \caption{
    Diagram (not to scale) of the angles formed between the three angular momentum vectors
    and a few of the key velocities in the problem.
    The grey band denotes the gas disc, and the red thin ellipse denotes the inner
    binary, which orbits about its centre of mass denoted by the blue dot.
    We denote the inner (red, $\bm{L}_{\rm in}$), outer (blue, $\bm{L}_{\rm
    out}$) and disc (black, $\bm{L}_{\rm disc}$) orbit normals with
    thin vectors, and we denote the outer orbital velocity (blue), disc Keplerian
    velocity (black), relative velocity $\bm{v}_{\rm out, rel}$ (green) upon
    disc passage with thick vectors.
    Finally, we illustrate that the inner orbital velocity lies in the plane of
    the orbit (along the thick orange vectors).
    The angle formed between the inner orbit normal and the outer orbit's
    relative velocity is $i_{\rm rel}$ (orange arc).
    }\label{fig:2diagram}
\end{figure}

The many sign changes of $\dot{a}$ in all of
Figs.~\ref{fig:results} and~\ref{fig:results2} and their dependence on the
orbital geometries can be understood.
We first consider two simple limiting cases: (i) when $\bm{v}_{\rm out, rel} \perp
\bm{v}$, inspiral occurs, while when instead (ii) $\uv{v}_{\rm out, rel}$ lies
in the plane of $\bm{v}$, then orbit averaging yields $\langle (\uv{v}_{\rm out,
rel} \cdot \bm{v})^2 \rangle = v^2/2$, resulting in outspiral.
As such, it is clear that in general the sign of $\dot{E}$ depends on the angle $\cos i_{\rm rel} \equiv \uv{v}_{\rm out, rel} \cdot \uv{l}$:
\begin{equation}
    \langle (\uv{v}_{\rm out, rel} \cdot \bm{v})^2 \rangle =
        \frac{v^2}{2}\sin^2 i_{\rm rel}.\label{eq:def_irel}
\end{equation}
The relative orientations of all the angles in the problem, including $i_{\rm
rel}$, is shown in Fig.~\ref{fig:2diagram}.
Outspiral is thus expected when $\sin^2 i_{\rm rel} > 2/3$, or when $i_{\rm rel} > 54^\circ$.
Notably, since $\bm{l} \perp \uv{z}$ for large $i_{\rm in}$ (as is
necessary for ZLK cycles), and $\uv{v}_{\rm out, rel}$ becomes
predominantly in the $\uv{z}$ direction for low $i_{\rm out}$, we find that
$\sin i_{\rm rel}$ is large when $i_{\rm out}$ is low.
\textit{This yields the intriguing conclusion that binaries tend to outspiral more
readily as their outer orbits become aligned with the disc.}
This effect is compounded by the fact that low-$i_{\rm out}$ orbits tend to have smaller $v_{\rm out, rel}$, so the drag felt by such binaries depends on the objects' gravitational cross sections.

However, from the lower left panels of Figs.~\ref{fig:results}
and~\ref{fig:results2}, it can be seen that the evolution of $a$ in the fiducial system is irregular
and is not easily described by any of the above results individually.
This is due to the close equality of $A_{\rm geom}$ and $A_{\rm grav}$ for our
fiducial parameters (Eq.~\ref{eq:crosssection_ratio}) as well as the natural
fluctuations introduced by the ZLK effect.
Nevertheless, all of the features observed in our simulations can be understood
in terms of the results above.
We identify three key ones below:
\begin{itemize}
    \item Note that systems prefer to inspiral at earlier times, and to
        outspiral at later times.
        This is because as the outer orbit aligns with the disc, the drag transitions to the BHL regime
        (Eq.~\ref{eq:crosssection_ratio}), while $i_{\rm rel} \to 90^\circ$ (Eq.~\ref{eq:def_irel}), leading to orbit softening.

    \item The system also fluctuates between in and outspiral on short
        timescales (evident in all of
        Fig.~\ref{fig:results_puffy} in the lower-left-most
        panels).
        We attribute this to the changing value of $i_{\rm rel}$ as ZLK
        oscillations change the inner orbit orientation.

    \item However, it is noteworthy that systems will sporadically transition
        from extended phases of outspiral back to extended periods of inspiral
        (most markedly at $\sim 15\;\mathrm{Myr}$ in the blue and green curves
        in Fig.~\ref{fig:results}).

        This can be understood by seeing that the $i_{\rm in}$ evolution also
        changes during these phases (see panels for $i$ in Fig.~\ref{fig:results}), where the maximum $i_{\rm in}$ becomes somewhat
        lower than $90^\circ$.
        This decreases $i_{\rm rel}$ (Eq.~\ref{eq:def_irel}), allowing the
        system to resume inspiraling despite being in the BHL regime.
        This qualitative change in the inclination oscillations
        arises because the ZLK cycle transitions from circulation $\omega$
        (argument of pericentre of the inner orbit) to libration
        \citep[e.g.,][]{katz2011, shevchenko2016lidov}, as can be seen in the bottom panels of Fig.~\ref{fig:results}.
\end{itemize}
In summary, the combination of the drag transitioning to the BHL regime, as well as the orientation of the inner orbit when experiencing the ZLK effect, causes binaries to widen as they align with the disc.
Such a widening weakens the short-range forces experienced by the orbit (Eqs.~\ref{eq:eps_gr}, \ref{eq:epsrot}), allowing the system to reach smaller pericentre distances.
This has important astrophysical applications as it can potentially induce stellar mergers, as can be seen in bottom-right panel of Fig.~\ref{fig:results} \citep[cf.][]{Stephan2016}.
We further discuss the consequences of this evolution in Section~\ref{sec:discussion}.

\begin{figure*}
    \centering
    \includegraphics[width=\textwidth]{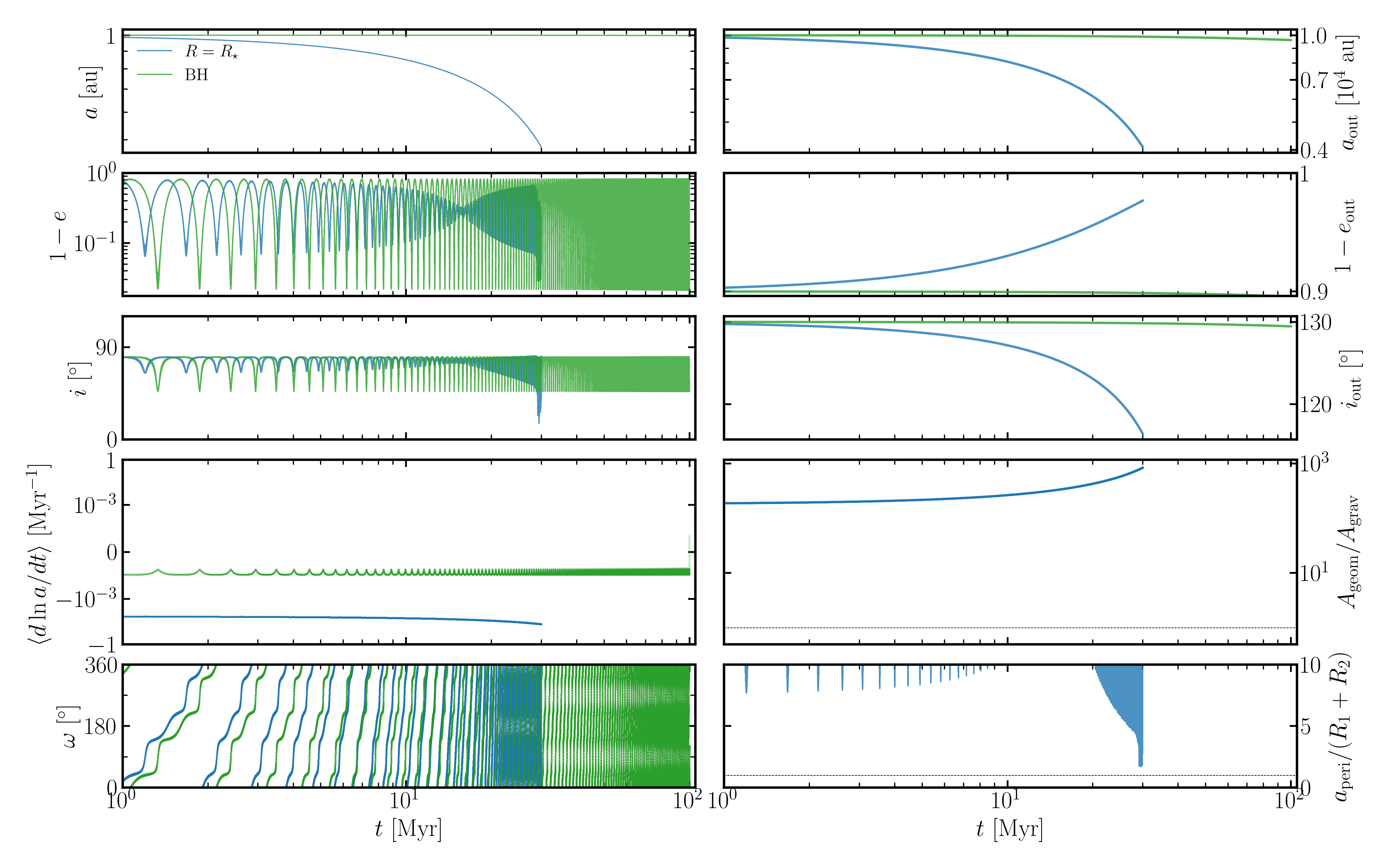}
    \caption{
    Same as Fig.~\ref{fig:results2} but with a larger initial $i_{\rm out} = 130^\circ$, and again comparing the evolution of a stellar binary to a compact object one.
    The dependence of the direction of evolution of the outer orbit's eccentricity $e_{\rm out}$ on the physics of aerodynamic drag is visible.
    }\label{fig:results_misal}
\end{figure*}

Finally, since orbits in a nuclear star cluster can be isotropically oriented, we also study the evolution of systems with significant misalignment of their outer orbits.
In Fig.~\ref{fig:results_misal}, we show the evolution of an orbit that
initially begins retrograde, $i_{\rm out} > 90^\circ$, for both a stellar and compact object binary.
We see that the stellar binary remains in the geometric drag regime (because the outer orbit decays and $v_{\rm rel}$ grows larger), resulting in consistent inner orbit hardening.
Additionally, the outer orbit's eccentricity and inclination both damp, but much more slowly than the orbit's semi-major axis.
This would result in the (inner) binary being disrupted by the SMBH.
On the other hand, the black hole binary experiences negligible orbital evolution, but undergoes gentle eccentricity pumping, as is expected from prior works \citep{secunda2021_retrograde, macleod2022_gasdf, wang2024_agn}.
We provide a brief description of the transition from eccentricity damping to pumping in Appendix~\ref{app:deoutdt}.

\subsection{Gas-ZLK interplay across the AGN parameter space}\label{ss:sim_realistic}

Finally, to illustrate the evolution of the inner binary for a few different parameters in realistic AGN discs, we present select integrations for the parameters indicated by the white crosses in Fig.~\ref{fig:connar}.
The three sets of orbital parameters we choose are motivated as follows:
In the rightmost cross, the parameters are quite similar to our fiducial parameters used in Fig.~\ref{fig:results}.
In the middle cross, the decreased values of both $a_{\rm out} = 3000\;\mathrm{au}$ and $a_{\rm in} \approx 0.25\;\mathrm{au}$ give $\epsilon_{\rm rot, 1} + \epsilon_{\rm rot, 2} \approx 0.4$, yielding a reduced value of $e_{\max} \approx 0.9$.
In the leftmost cross, the further decreased values of $a_{\rm out} = 400\;\mathrm{au}$ and $a_{\rm in} \approx 0.04\;\mathrm{au}$ yield $\epsilon_{\rm rot, 1} + \epsilon_{\rm rot, 2} \approx 11$, leading to complete suppression of ZLK eccentricity oscillations.
Across these three sets of parameters, the gas density $\rho_g$ increases as $a_{\rm out}$ decreases ($\approx [10^{10}, 4 \times 10^{11}, 10^{13}]\;M_\odot / \mathrm{pc}^3$ for the three parameters), leading to more rapid evolution of the system architecture.

\begin{figure*}
    \centering
    \includegraphics[width=\textwidth]{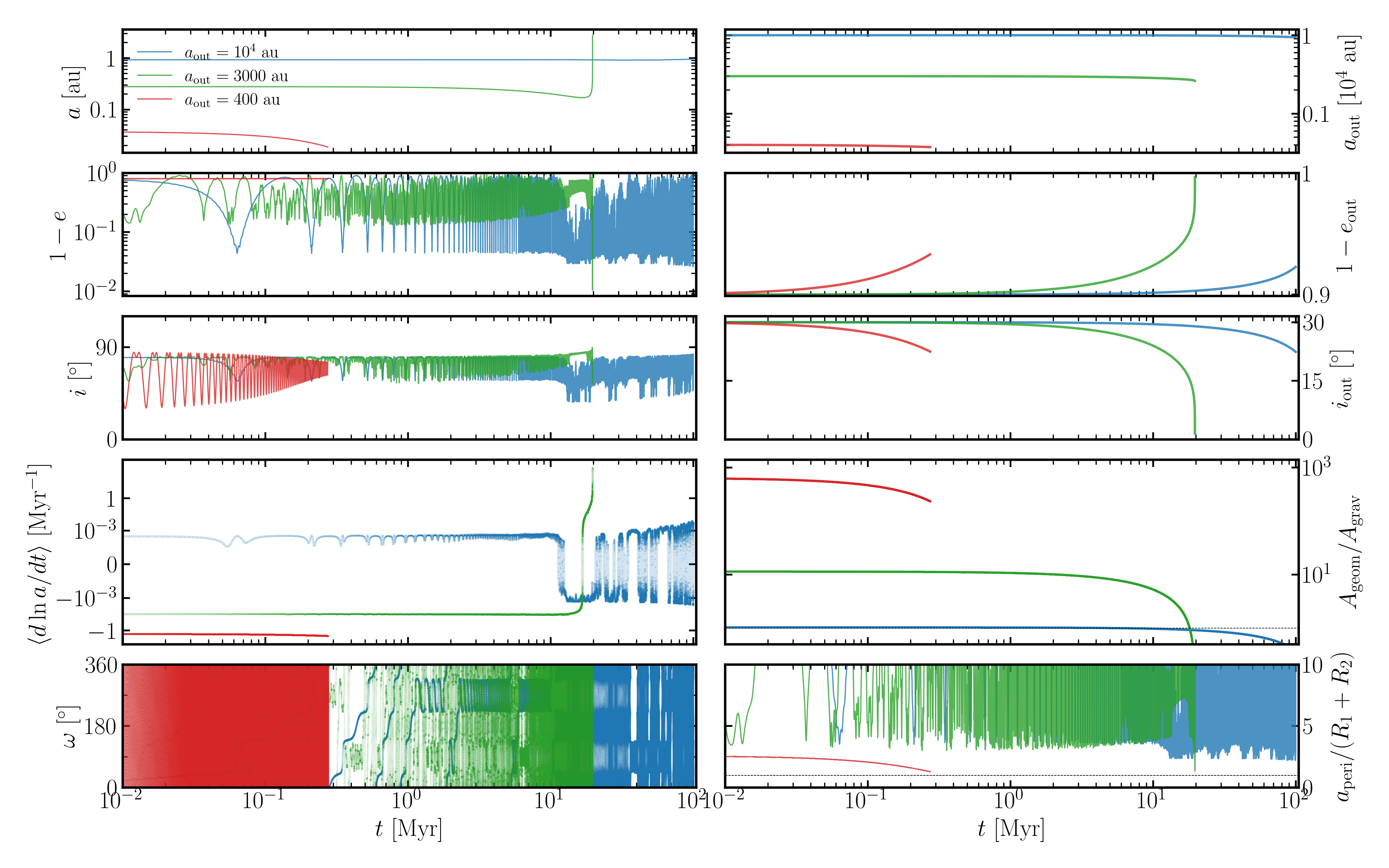}
    \caption{The evolution of systems with parameters corresponding to the three
    crosses shown in Fig.~\ref{fig:connar}. It can be seen that the maximum
    eccentricities become smaller for more compact binaries. The orbital
    evolution timescale decreases as $a_{\rm out}$ decreases, since the gas
    density increases.}
    \label{fig:connar_examples}
\end{figure*}

The evolutions for these three sets of parameters are shown in Fig.~\ref{fig:connar_examples}.
As expected, the system's maximum eccentricity decreases as $a_{\rm out}$ decreases, but the evolution of both $a_{\rm in}$ and $a_{\rm out}$ speeds up due to the increased gas density: note that the outer orbit's evolutionary timescale as given by Eq.~\eqref{eq:t_outdrag} is $\sim [150, 3, 0.04]\;\mathrm{Myr}$ for the three parameters, in good agreement with the simulation results.
Note that in the most compact case ($a_{\rm out} = 400\;\mathrm{au}$), the oscillation of $e_{\rm in}$ due to the ZLK effect is completely suppressed as expected;
the oscillations in $i$, the misalignment angle between the inner orbit and the \textit{disc}, are due to the rapid precession of the inner orbit about the \textit{outer orbit}, which itself is misaligned from the disc.

\section{Discussion}
\label{sec:discussion}

In this section, we begin by highlighting a few key dynamical results of interest, then we identify interesting astrophysical applications of our results, and we lastly discuss caveats and directions for future work.

\subsection{Dynamical Implications}

First, note that, depending on whether the stars experience geometric or BHL drag, and depending on the orbital geometry, the inner orbit can either harden (geometric, or BHL in the limit $\sin^2 i_{\rm rel} < 2/3$, Eq.~\ref{eq:def_irel}) or soften (BHL in the limit $\sin^2 i_{\rm rel} > 2/3$).
In general, since the orientation of the inner binary can fluctuate due to ZLK oscillations induced by the central SMBH, the evolution of the inner orbital separation is often erratic.

Second, because binaries on low-$i_{\rm out}$ orbits penetrate the disc with small orbital velocities, a binary will tend to transition to the BHL-dominated regime as it is driven towards alignment with the AGN disc (Eq.~\ref{eq:crosssection_ratio}).
As alignment occurs, $i_{\rm rel}$ generally becomes large as well (Section~\ref{ss:ain_dot}).
As such, the orbits of binaries tend to \textit{soften} as the inclination of their orbit around the central SMBH decreases.
Since the inner and outer orbits tend to evolve on comparable timescales (Eqs.~\ref{eq:t_outdrag},\ref{eq:t_indrag}), the inner orbit readily survives for a lifetime comparable to the alignment time of the outer orbit.

\subsection{Applications to Other Systems}

Secular dynamics in gas-rich environments have various astrophysical implications, that share common principles with the system we described here. Here we will briefly review some of them.

The growth we described above of $a_{\rm in}$ is accompanied by a weakening of the ZLK short-range forces (Section~\ref{ss:zlk}).
This allows for the inner binary's pericentre separation to decrease as its orbit expands, potentially leading to stellar collisions once $a_{\rm in, peri} \sim R_1 + R_2$.
As studies suggest that blue stragglers may be the result of dynamically-driven stellar mergers \citep{perets2009_bluestragglers, naoz2014_mergers, Stephan2016}, the mechanism studied here may suggest an enhanced formation rate of such merger products.
This mechanism may also increase the merger rate of binary black holes around SMBHs, by increasing the inner orbital separation, playing a similar role to weak flybys \citep[e.g.,][]{Winter2024}.

Another gas-rich environment that could host hierarchical triple systems is gas-rich clusters. While the gas density in clusters today is negligible, clusters are thought to host significant gas fractions during the formation of second (or further) stellar generations, or due to accretion from external sources in general \citep[see][for a detailed review]{BastianLardo2018}.
Nuclear star clusters (NSCs) are one of the densest environments in the universe, hosting millions of stars within a radius of a few parsecs. Various models suggested that these clusters were gas-rich environments for extended epochs, either because gas fueling towards the centres of galaxies during early stages or later accretion \citep[e.g.,][]{Loose1982,Milosavljevic2004,Guillard2016}. Also there, all the binaries could be considered as triples with the central supermassive black hole (SMBH) as the third object.
It should be noted that the geometry of these systems is inherently different from the disc-crossing we discussed here, and accordingly, the problem should be treated with the necessary modifications. The gas effect will be more prominent here, as all the objects are embedded in the gaseous medium during their motion.

For the systems we consider with larger misalignments with respect to the AGN disc (Fig.~\ref{fig:results_misal}), we find that the outer orbits of stellar binaries can decay relatively quickly (blue), while the orbits of black hole binaries do not (green).
It is therefore tempting to conclude that binary dissociation and formation of hypervelocity stars via the Hills mechanism \citep{hills1988, brown2015_hvs, han2025_hvslmc} can arise from inward migration of stellar binaries.
However, since the inner orbit coalesces at a similar rate (as seen in the top-left panel of Fig.~\ref{fig:results_misal} and predicted by Eqs.~\ref{eq:t_outdrag} and~\ref{eq:t_indrag}), the dynamical stability of the binary does not change dramatically (Eq.~\ref{eq:mardling}), and no hypervelocity star is produced.

The combined effect of the ZLK mechanism and relaxation (either scalar or vector) could play an important role in catalyzing gravitational wave mergers \citep[e.g.,][]{Naoz2022}. Introducing a gas-rich background to this system will modify the dynamics of this system as well, potentially catalyzing the mergers even further.

\subsection{Caveats \& Future directions}

In this section, we briefly discuss some additional physical effects neglected in this work, their impacts, and potential avenues for future exploration.

First, the interaction of objects with gas-rich environments is complex and not well understood.
For simplicity, we have adopted here the simple prescription accounting for the physical and accreting radii of the binary components (Eq.~\ref{eq:def_Fdrag}).
Most notably, we have neglected the effect of gas dynamical friction \citep{Ostriker1999}.
However, because the physical scalings for dynamical friction are the same as that of BHL accretion-driven drag \citep[e.g.,][]{wang2024_agn}, the inclusion of gas dynamical friction is not expected to change our key results concerning the boundary between the hardening and softening of the inner binary, only the rate at which it occurs.
Notably, such a transition also appears when explicitly modeling the competition between gas dynamical friction and accretion \citep{GrishinPeretsII2016}, and the analytical results presented in this paper can in principle be extended to predict the transition between in- and out-spiral in the presence of gas dynamical friction as well.
The gas can also directly affect the binary components when there is significant mass accretion \citep{spieksma2025_accretion}.
In addition, the binaries considered here are soft: their orbital velocities are smaller than the typical velocity dispersion of other stars in the nuclear star cluster.
As such, two-body relaxation will also act to gradually soften the binary \citep{heggie1975_law, collins2008_levyflight, Winter2024}.

It should be noted that we've treated the disc as approximately homogeneous throughout the disc passage (assuming that each mass ballistically passes through the disc), but the disc passage time can easily be much longer than $P_{\rm in}$.
Previous results in the literature suggest that the drag experienced by the binary components should be independent when $a_{\rm in} \gg Gm_{12} / v_{\rm rel}^2$ \citep[e.g.,][]{comerford2019_bhlregimes}, which is satisfied by our fiducial parameters (Eq.~\ref{eq:crosssection_ratio}).
When this relation is not satisfied, the wakes generated by the binary components can interact, affecting the overall drag on the binary; we defer the hydrodynamic exploration of this regime to future work.
We anticipate that the inner orbital evolution is likely to change, as each mass individually experiences very different drag forces, but the outer orbital evolution will likely be qualitatively similar.

Our finding that binaries tend to transition from orbital hardening to softening
as their outer orbits align with the disc must be qualified:
as the inclination approaches $h \equiv H/r$, some of the approximations made break
down: the gas can follow the binary as $v_{\rm rel} \to c_s$, invalidating the
$\sim$ ballistic approximation we've implicitly assumed by treating the disc as
a homogeneous medium with density $\rho_g$.
Additionally, $\bm{v}_{\rm out}$ will no longer be approximately constant as the
arclength of the orbit within the disc becomes non-negligible.

We note a few higher-order ZLK effects we've neglected for simplicity and because they are expected not to affect the qualitative evolution of the systems considered here.
First, the octupole-order ZLK effect \citep{ford2000_octlk, katz2011, Naoz2016}
is small as long as $\epsilon_{\rm oct} \ll 1$, where
\begin{align}
    \epsilon_{\rm oct}
        &\equiv \frac{m_1 - m_2}{m_{1} + m_2}
            \frac{a}{a_{\rm out}}
            \frac{e_{\rm out}}{1 - e_{\rm out}^2}.
\end{align}
While both the first and third terms can be of order unity, the small magnitude of the second term $a / a_{\rm out} \sim \mathcal{O}(10^{-4})$ ensures that this is negligible.
Second, the corrections to the double-averaged equations of motion are negligible when $\epsilon_{\rm SA}$ is small \citep{luo2016, grishin2018, tremaine2023_brown}, where
\begin{align}
    \epsilon_{\rm SA}
        &\equiv
            \p{\frac{m_3^2}{m_{12}m_{123}}}^{1/2}
            \p{\frac{a}{a_{\rm out}j_{\rm out}^2}}^{3/2},\\
% >>> (1e7 Msun) / ((3 Msun) * (1e7 Msun))^(1/2) * (1 / 1e4)^(3/2)
% 0.001826
        &\sim
            10^{-3}
            \p{\frac{m_3}{m_{12}}}^{1/2}
            \p{\frac{a / (a_{\rm out} j_{\rm out}^2)}{10^{-4}}}^{3/2}.
\end{align}
Finally, the back-reaction on the outer orbit \citep{LL18, mangipudi2022} is negligible due to the significantly larger angular momentum of the outer orbit.

In calculating the drag force (Eq.~\ref{eq:def_Fdrag}), we have neglected the
additional component of the vertical velocity due to the disc gravity.
Using the results in Section~\ref{ss:diskp_eom} (and Appendix~\ref{app:prec}),
it is easy to show that
\begin{align}
% >>> (2 * pi * G * (1e11 Msun) / (1 pc)^3 * 0.005 * (1e4 AU) * (1e4 AU))^(1/2)
% 1.782410e+05 m\s
    v_{z, \mathrm{disc}}
        &\sim \sqrt{2\pi G \Sigma z}\nonumber\\
        &\simeq
            200\sqrt{\sin i_{\rm out}}\;\mathrm{km/s}
            \p{\frac{\rho_g}{10^{11}M_{\odot}/\;\mathrm{pc}^3}}^{1/2}
            \p{\frac{a_{\rm out}}{10^4\;\mathrm{au}}}^{-1}.
\end{align}
Recall that the vertical component of $\bm{v}_{\rm out, rel}$ is given by
\begin{equation}
% >>> (G * (1e7 Msun) / (1e4 AU))^(1/2)
% 9.419865e+05 m\s
    v_{\rm out}\sin i_{\rm out}
        \simeq
            10^3\sin i_{\rm out}\;\mathrm{km/s}
            \p{\frac{m_3}{10^7 M_{\odot}}}^{1/2}
            \p{\frac{a_{\rm out}}{10^4\;\mathrm{au}}}^{-1/2}.
\end{equation}
Thus, we see that the contribution from the disc's gravity to the particle's
vertical velocity is small if $i_{\rm out} \gtrsim 1/25 \approx 3^\circ$.
Thus, our inclination damping rate is moderately underestimated by neglecting
this contribution.
However, most of the inaccuracy arises when the inclination becomes small, where
inclination damping is already quite efficient, so the quantitative effect on
our results may be modest.

Finally, we have neglected the interactions between the components of the inner binary, including the enhanced dissipation that can arise as the inner pericentre distance decreases.
The most relevant such effect in stellar binaries is tidal dissipation \citep{FabryckyTremain2007}.
Dissipation may affect our conclusion that the binaries we study can sometimes be driven towards stellar merger, as the small pericentre distances may result in sufficient dissipation to overwhelm the drag-driven orbit softening.
Analogous conclusions can be drawn for the emission of gravitational waves by binary compact objects.
We defer the quantitative exploration of this competition to future work.

In accretion discs, gas exhibits differential shear motion, which deviates from Keplerian motion. In this paper, we ignore this effect and assume Keplerian velocities. Including corrections due to shearing motion could add corrections to the results derived above, and we leave this for future work.

\section{Summary}
\label{sec:summary}
In this paper, we studied the evolution of hierarchical triples in gas-rich AGN discs. The system we studied consists of a central SMBH, and a binary revolving around it, crossing a gas-rich disc twice in a period.
All the binaries in an accretion disc should be considered as a hierarchical triple with the central object as a third perturber, and hence the problem we investigate is general and relates to any gas-rich accretion disc.

One of our principal results concerns the evolution of the orbital separation of the inner binary.
Our linear analysis shows that orbital expansion is possible when two conditions are satisfied: (i) the angle between the binary's orbit normal and its relative velocity to the disc fluid exceeds $\sim 54^\circ$, and (ii) the binary components' gravitational/Bondi-Hoyle-Lyttleton cross sections dominate their physical cross sections (see Section~\ref{ss:ain_dot}).
These two conditions are easily met by binaries experiencing ZLK oscillations driven by the central SMBH as they gradually become aligned with the AGN disc due to gas drag.
As these orbits expand, their minimum pericentre distances decrease.
This effect can drive enhanced production of stellar mergers and gravitational wave sources.

Secular dynamics in gas-rich environments are essentially different, and lead to various phenomena, including enhancing the rates of gravitational waves, and other transients.
An analogous treatment could be applied to gas-rich spherical configurations, with the necessary modifications, such as triples embedded in gas-rich nuclear star clusters.

\section*{Acknowledgements}

We thank the anonymous referee for helpful comments that improved the quality of this paper.
We thank Mark Dodici, Jeremy Goodman, Melaine Saillenfest, and Scott Tremaine for helpful comments and suggestions that improved the quality of this manuscript.
This work was inspired by conversations at the New Ideas on the Origin of Black Hole Mergers workshop at the Niels Bohr Institute.
YS is supported by a Lyman Spitzer, Jr. Postdoctoral Fellowship at Princeton
University.
The research leading to this work was supported by the Independent Research Fund Denmark via grant ID 10.46540/3103-00205B.

%%%%%%%%%%%%%%%%%%%%%%%%%%%%%%%%%%%%%%%%%%%%%%%%%%
\section*{Data Availability}

The data underlying this article will be shared on reasonable request to the corresponding author.

%%%%%%%%%%%%%%%%%%%% REFERENCES %%%%%%%%%%%%%%%%%%

% The best way to enter references is to use BibTeX:

\bibliographystyle{mnras}
\bibliography{example} % if your bibtex file is called example.bib

% Alternatively you could enter them by hand, like this:
% This method is tedious and prone to error if you have lots of references
%\begin{thebibliography}{99}
%\bibitem[\protect\citeauthoryear{Author}{2012}]{Author2012}
%Author A.~N., 2013, Journal of Improbable Astronomy, 1, 1
%\bibitem[\protect\citeauthoryear{Others}{2013}]{Others2013}
%Others S., 2012, Journal of Interesting Stuff, 17, 198
%\end{thebibliography}

%%%%%%%%%%%%%%%%%%%%%%%%%%%%%%%%%%%%%%%%%%%%%%%%%%

%%%%%%%%%%%%%%%%% APPENDICES %%%%%%%%%%%%%%%%%%%%%

\appendix

\section{Equations of Motion}\label{app:eom}

We provide here a complete reference of our implementation of the secular equations of motion (giving rise to the ZLK effect) and of the gas drag experienced by the binary as it passes through the AGN disc.

\subsection{Secular Equations}\label{app:eom_zlk}

In this paper, we adopt the standard double-averaged quadrupole-order equations for the evolution of the inner and outer orbits of the binary \citep[e.g.,][]{liu2015suppression}
\begin{align}
    \frac{d\bm{L}}{dt}
        ={}& \frac{3L_0}{4 t_{\rm ZLK}}
            \Big[
                (\bm{j} \cdot \uv{l}_{\rm out})(\bm{j} \times \uv{l}_{\rm out})
            - 5 (\bm{e} \cdot \uv{l}_{\rm out})(\bm{e} \times \uv{l}_{\rm out})
            \Big],\label{eq:ZLK1}\\
    \frac{d\bm{e}}{dt}
        ={}&
            \frac{3}{4 t_{\rm ZLK}}
            \Big[
                (\bm{j} \cdot \uv{l}_{\rm out})(\bm{e} \times \uv{l}_{\rm out})
                + 2 \bm{j} \times \bm{e}\nonumber\\
            &- 5 (\bm{e} \cdot \uv{l}_{\rm out})(\bm{j} \times \uv{l}_{\rm out})
            \Big],\label{eq:ZLK2}\\
    L_0
        \equiv{}&
            \mu \sqrt{G m_{12} a},\\
    \frac{1}{t_{\rm ZLK}}
        ={}&
            \frac{m_3}{m_{12}}\left(\frac{a}{\tilde{a}_{\rm out}}\right)^3
            n,\label{eq:t_ZLK}
\end{align}
where $\bm{j}=\sqrt{1-e^2}\uv{l}_{\rm out}$ is the dimensionless angular momentum vector, $t_\text{ZLK}$ is the ZLK timescale, $\tilde{a}_{\rm out} \equiv a_{\rm out}\sqrt{1 - e_{\rm out}^2}$, $m_{12} = m_1 + m_2$, $\mu = m_1m_2/m_{12}$, and $n \equiv \sqrt{Gm_{12} / a^3}$ is the inner orbit's mean motion.

In addition to these equations, we include three sources of apsidal precession that can affect the inner orbit's evolution if the pericentre distance is sufficiently small: first-order post-Newtonian general relativistic (GR) effects, and the rotational and tidal bulges of both stars.
These are given by \citep{Kiseleva1998, FabryckyTremain2007, liu2015suppression}
\begin{align}
    \frac{d\bm{e}}{dt}
        ={}&
           (\Omega_{\rm GR} + \Omega_{\rm Tide} + \Omega_{\rm Rot})
           (\uv{l} \times \bm{e}),\\
    \Omega_{\rm GR}
        ={}&
            \frac{3G^2m_{12}^2 \mu}
                {c^2a^2 L},\\
    \Omega_{\rm Tide}
        ={}&
            \frac{15}{(1 - e^2)^5}
            \p{1 + \frac{3e^2}{2} + \frac{e^4}{8}}\nonumber\\
        &\times\left[\frac{k_{2,1}m_1}{m_2}\p{\frac{R_1}{a}}^5
                + \frac{k_{2,2}m_2}{m_1}\p{\frac{R_2}{a}}^5\right]n,\\
    \Omega_{\rm Rot}
        ={}&
            \frac{k_{2,1} m_1\omega_1^2R_1^5
                + k_{2, 2}m_2\omega_2^2R_2^5}{
                2Ga^2m_1m_2(1 - e^2)^2}n,
\end{align}
where $k_{2, i}$ denotes the tidal Love number of the stars (we take this to be $\sim 0.04$ for both stars, the solar value, \citealp{mecheri2004_sunJ2})
% >>> 2.2e-7 / ((2 * pi / (27 day))^2 / (G * Msun / Rsun^3))
% 0.011966
, and $\omega_i$ their spin frequencies.

\subsection{Gas Drag}\label{app:eom_disc}

Here, we develop convenient expressions for the drag experienced by a binary penetrating the AGN disc twice per outer orbital period.
By avoiding direct reference to orbital elements, our equations are easy to implement alongside those given above in Appendix~\ref{app:eom_zlk}.

As discussed in the main text, the inner and outer orbits evolve due to the drag force on the relative separation $\bm{F}_{\rm d, rel}$ and $\bm{F}_{\rm d, com}$ (Eqs.~\ref{eq:F_d_decomposition}).
These forces then modify inner and outer orbits' angular momentum and eccentricity vectors
\citep[e.g.,][]{heggie1996_rasio} on each orbit
crossing via
\begin{align}
    \frac{d\bm{L}}{dt}\bigg{|}_\text{gas}
        \simeq{}& \bm{r} \times \bm{F}_{\rm d} \frac{t_{\rm cross}}{P_{\rm out}},\\
    \frac{d\bm{e}}{dt}\bigg{|}_\text{gas}
        \simeq{}&
            \frac{t_{\rm cross} / P_{\rm out}}{Gm_{12}}\Big(
                2(\bm{F}_{\rm d} \cdot \bm{v})\bm{r}
                - (\bm{r} \cdot \bm{v})\bm{F}_{\rm d}
                - (\bm{F}_{\rm d} \cdot \bm{r})\bm{v}
            \Big).\label{eq:dedt_hr96}
\end{align}
Here, $t_{\rm cross} = 2H(r_{\rm out, cross}) / |v_{\rm out} \sin i_{\rm out}|$ is the time it takes the binary to cross the disc at distance $r_{\rm out, cross}$,
and we adopt $H(r_{\rm out}) / r_{\rm out} = 1 / 200$.
Summing over the two disc passages and their associated $\bm{r}_{\rm i}$ and $\bm{v}_{\rm rel, i}$, we can solve for the long-term evolution of the inner and outer orbits simultaneously.

In order for the above gas-driven evolution, which depends on the state of the orbits at specific orientations and instances in time, to be easily integrated in conjunction with the vector-form ZLK equations (Eqs.~\ref{eq:ZLK1} and~\ref{eq:ZLK2}), it is helpful to re-express the orbital positions and velocities in terms of the orbital eccentricity and angular momentum vectors.
First, to evaluate the inner orbit's configuration, it is easiest to adopt the approach of Gauss's $f$ and $g$ functions \citep{tremaine_book} to transform the state of the inner orbit at pericentre to an arbitrary true anomaly $f$:
\begin{align}
    \bm{r}_{\rm peri}
        &= \frac{L^2}{\mu^2 Gm_{12}(1 + e)}\uv{e},\\
    \bm{v}_{\rm peri}
        &= n\uv{l} \times \bm{r}_{\rm peri} \frac{\sqrt{1 + e}}{(1 - e)^{3/2}},\\
    \bm{r}
        &=
            \frac{\cos f + e \cos f}{1 + e \cos f}\bm{r}_{\rm peri}
            + \frac{(1 - e^2)^{3/2}\sin f}{n(1 + e\cos f)(1 + e)}\bm{v}_{\rm peri},
            \label{eq:r_gauss}\\
    \bm{v}
        &=
            -n\frac{e\sin f + \sin f}{(1 - e^2)^{3/2}}\bm{r}_{\rm peri}
            + \frac{e + \cos f}{1 + e}\bm{v}_{\rm peri}.
\end{align}
Note that Eq.~\eqref{eq:r_gauss} reduces to the well-known $\bm{r} / r = \cos f
\uv{r}_{\rm peri} + \sin f (\uv{l} \times \uv{r}_{\rm peri})$
\citep{murray_dermott_1999, tremaine_book}.

On the other hand, for the outer orbit, standard celestial mechanics identities can be used to show that \citep{tremaine_book}
\begin{align}
    \uv{r}_{\rm out, cross, \pm}
        &\propto \pm \uv{z} \times \uv{l}_{\rm out},\\
    r_{\rm out, cross, \pm}
        &= \frac{a_{\rm out}(1 - e_{\rm out}^2)}{1 + \bm{e}_{\rm out} \cdot \uv{r}_{\rm out, cross, \pm}},\\
    \bm{v}_{\rm out, cross, \pm}
        &= \sqrt{\frac{Gm_{123}}{a_{\rm out}(1 - e_{\rm out}^2)}}
            \uv{l}_{\rm out} \times \p{\uv{r}_{\rm out, cross, \pm} + \bm{e}_{\rm out}},\\
    \bm{v}_{\rm disc}
        &=
            \sqrt{\frac{Gm_{12}}{r_{\rm out, cross, \pm}}}
            \uv{z} \times \uv{r}_{\rm out, cross, \pm}.
\end{align}

\subsection{Precession due to Disc}\label{app:prec}

In this appendix, we provide a simple derivation for the nodal precession rate of a particle with semi-major axis $a$ due to a razor-thin disc with inner and outer radii $r_{\rm in} < a < r_{\rm out}$.
As discussed in the main text, standard secular approaches are inapplicable due to the divergence when $a = r$.
Instead, we estimate the precession frequency by computing the various epicyclic frequencies introduced by the disc's gravitational potential: recall that the closed nature of the Keplerian orbit is due to the equality of the radial and vertical epicyclic frequencies with the orbital frequency, and deviations give rise to precession.
For simplicity, we specialize to a $p=-1$ surface density profile ($\Sigma = \Sigma_0 a_0 /a$), which is known as a Mestel disc \citep{binneytremaine_book}, and focus on circular orbits.
The potential of such a thin disc can be analytically expressed in cylindrical coordinates as \citep{binneytremaine_book}
\begin{align}
    \Phi(a, z)
        &=
            \int\limits_0^\infty
                S(k) J_0(ka)e^{-k|z|}\;\mathrm{d}k,\\
    S(k)
        &=
            -\frac{2\pi G \Sigma_0 a_0}{k},
\end{align}
where $J_0$ is a Bessel function of the first kind.
The resulting vertical acceleration at height $z$ due to the disk is given by
\begin{align}
    \ddot{z}_{\rm disk}
        &=
            -\frac{\partial \Phi}{\partial z}\nonumber\\
        &=
            -2\pi G\Sigma_0a_0 \sgn(z)
                \int\limits_0^\infty
                    J_0(ka) e^{-kz}\;\mathrm{d}k\nonumber\\
        &=
            -2\pi G\frac{\Sigma_0a_0}{a} \sgn(z)
            \frac{1}{(1 + (z/a)^2)^{1/2}}\nonumber\\
        &=
            -2\pi G\frac{\Sigma_0a_0}{a} \sgn(z)
            ,\label{eq:zddot}
\end{align}
where we have used the identity \citep{binneytremaine_book}
\begin{align}
    \int\limits_0^\infty
        e^{-\alpha t}J_\nu(t)t^{\nu + 1}\;\mathrm{d}t
            = \frac{2^{\nu + 1}\p{\nu + \frac{1}{2}}!}{\sqrt{\pi}
                \p{1 + \alpha^2}^{\nu + 3/2}},
\end{align}
for $\nu = 0$ and $\alpha = (z/a)$.

This additional anharmonic vertical acceleration adds to the Keplerian vertical restoring force.
The resulting vertical epicyclic frequency is given by seeking solutions of form $z = A\sin(\omega t)$ to the differential equation
\begin{equation}
    \ddot{z} + n^2z
        = -2\pi G\frac{\Sigma_0a_0}{a} \sgn(z).
\end{equation}
The dominant oscillation frequency ($\omega \approx n$, as long as $G \Sigma_0 a_0 / a \ll n$) can be used to compute the precession frequency $\Omega_{\rm out} \equiv n - \omega$, resulting in
\begin{align}
    \Omega_{\rm out}
        &\simeq
            -\frac{4G}{n z_0}\frac{\Sigma_0 a_0}{a}\nonumber\\
        &=
            -\frac{G\rho_{g}}{n}\frac{4h}{\sin i},\label{eq:app_diskprec}
\end{align}
where $h \equiv H/r$, and the negative sign indicates retrograde precession.
Note that $\Omega_{\rm out}$ diverges as $i \to 0$, since $\ddot{z}_{\rm disk} \neq 0$ as $z \to 0$.
Physically, this is regularized when $i \lesssim h$, where the orbit is embedded in the the disc.
Since we do not consider low-inclination orbits in this paper, Eq.~\eqref{eq:app_diskprec} is appropriate.
Nevertheless, we note that this closely resembles the standard expression of $\Omega_{\rm out} \simeq \pi G\rho_{g} / n$ for orbits embedded in the disk (i.e.,\ $\sin i < h$; \citealp{ward1981, hahn2003, zanazzilai2018}), suggesting a natural transition between the two regimes.
% In practice, we adopt a constant, effective $\Omega_{\rm out, eff}$ satisfying $\Omega_{\rm out} \ll \Omega_{\rm out, eff} \ll 2\pi / t_{\rm ZLK}$, which gives the qualitatively correct evolution without needing to resolve the rapid orbital precession.

Finally, note that while the above calculation gives the nodal precession rate, $\bm{e}_{\rm out}$ will also evolve due to apsidal precession.
A similar calculation could be repeated for such purposes.
However, since $\bm{e}_{\rm out}$ nearly does not affect the dynamics (having only a small effect on the outer orbital velocities at each disc passage), we neglect the disc-driven apsidal precession of the outer orbit for simplicity.

\section{Eccentricity Evolution of Outer Binary}\label{app:deoutdt}

The eccentricity evolution of the outer binary is in general a function of the surface density profile of the disc, the drag prescription, and the argument of periastron of the outer orbit ($\omega_{\rm out}$).
Results have been derived for specific surface density profiles \citep[e.g.,][]{wang2024_agn} and for a static gas background \citep{macleod2022_gasdf}.
Here, we provide a simple derivation of the boundary between eccentricity pumping and damping.
For simplicity, we assume just a single object (and drop the `out' subscripts).
For now, we take $\omega = 0$; eccentricities tend to favor damping as $\omega$ is taken to be further from $0$ and $\pi$ \citep{wang2024_agn}, so our result should be taken as a necessary but not sufficient requirement for eccentricity pumping.

By focusing on the $\omega=0$ case, the two disc passages of the object are at its pericentre and apocentre.
At these points, $\bm{r} \perp \bm{v}$, and so Eq.~\eqref{eq:dedt_hr96} reduces to
\begin{equation}
    \delta e
        \propto
            t_{\rm cross}
            (\bm{F}_{\rm d} \cdot \bm{v}) \bm{r}.
            \label{eq:deltae_app}
\end{equation}
Here, we are only concerned with the relative magnitudes of $\delta e$ at pericentre (which leads to eccentricity damping, by taking the dot product of $\bm{e}$ with Eq.~\ref{eq:deltae_app}) and apocentre (which leads to eccentricity pumping).
By defining $\cos \theta = \uv{F}_{\rm d} \cdot \uv{v}$ and noting that $rv = L/\mu$ is a constant, we can rewrite
\begin{equation}
    \delta e
        \propto
            \frac{H}{v|\sin i|}
            F_{\rm d} \cos \theta.
\end{equation}
Defining
\begin{equation}
    F_{\rm d} \propto \rho v_{\rm rel}^\alpha,
\end{equation}
where $\alpha = 2$ is the geometric regime, and $\alpha = -2$ the BHL one, we obtain
\begin{equation}
    \delta e
        \propto
            \Sigma r
            v_{\rm rel}^\alpha\cos \theta.
\end{equation}
Now, by applying standard trigonometric results to the vectors shown in Fig.~\ref{fig:2diagram}, we find that
\begin{align}
    v
        &= v_{\rm disc} \cos i
            - v_{\rm rel} \cos \theta,\\
    \delta e
        &\propto
            -\Sigma r
            v_{\rm rel}^{\alpha - 1}
            (v - v_{\rm disc}\cos i).
\end{align}
Another application of the law of cosines shows that
\begin{equation}
    v_{\rm rel}^2
        =
            v^2
            + v_{\rm disc}^2
            -2vv_{\rm disc}\cos i.
\end{equation}
Finally, recall that, at pericentre and apocentre, $v_{\rm p, a} \propto \sqrt{(1 \pm e) / (1 \mp e)}$, $r_{\rm p, a} \propto 1 \mp e$, and $(v_{\rm disc})_{\rm p, a} \propto \sqrt{1 / (1 \mp e)}$.
With these expressions, and defining as before $\Sigma \propto r^{-p}$, we can expand in the limit $e \ll 1$ to find
\begin{equation}
    (\delta e)_{\rm p}
    + (\delta e)_{\rm a}
        \propto
            -e \cos i
            - 2e(1 - \cos i)
            \p{p + \frac{3(\alpha - 1)}{4}}
            + \mathcal{O}(e^2).
\end{equation}
Note that the prefactors we have dropped are positive.

For a Mestel disc ($p = 1$ as used in the text), we find that geometric drag ($\alpha = 2$) always corresponds to eccentricity damping, while BHL drag ($\alpha = -2$) gives eccentricity pumping when $i > 44^\circ$.
For a standard gas pressure-supported AGN disc, with $p = 3/5$ \citep{sirko2003_agn}, we obtain guaranteed eccentricity damping for geometric drag and eccentricity pumping for BHL drag when $i > 40^\circ$.

Again, we caution that these results are for $\omega=0$, and averaging over $\omega$ (as would be necessary when disc-driven apsidal precession is accounted for) will further prefer eccentricity damping \citep[as studied in][]{wang2024_agn}, increasing the critical inclination required for eccentricity growth.
Finally, note that evolution of the disc surface density can also increase $e_{\rm out}$ \citep{kaur2025_evolvingdisk_eout}.

% Don't change these lines
\bsp	% typesetting comment
\label{lastpage}
\end{document}